# SELF-CONSISTENT EQUATIONS FOR NONEMPIRICAL TIGHT BINDING THEORY

Alexander V. Mironenko*
*Department of Chemical and Biomolecular Engineering,*
*University of Illinois Urbana-Champaign, Urbana, Illinois 61820*

**Email: alexmir@illinois.edu*

A new reference state for density functional theory, termed the independent atom ansatz, is introduced in this work. This ansatz allows for the exact representation of electron density in terms of non-interacting, atom-localized orbitals. Self-consistent equations for localized states are derived. Total energy functional is found to closely resemble tight binding theory. The independent atom ansatz facilitates partial cancellation of inter-atomic electron-electron and electron-nuclear interactions, which allows for the derivation of analytical Hamiltonian matrix elements in a weak interaction limit. The formalism provides charge and energy decomposition analyses at no additional cost. It also includes mechanisms to remove self-interaction and static correlation errors. Initial numerical results for simple model systems have been previously reported [Mironenko, J. Phys. Chem. A **127**, 7836 (2023)].

## I. INTRODUCTION

Delocalized electronic states are fundamental to modern electronic structure methods.[1-4] While they are substantiated by experimental electronic spectra[5, 6] and allow for chemically accurate *in silico* predictions of molecular[7, 8] and materials'[9, 10] properties, the existence of delocalized states stands in contrast to the perceived locality of chemical interactions, attributed to the "nearsightedness of electronic matter".[11] Historically, the idea of a functional group localized within a molecule explained empirical chemical reactivity patterns,[12, 13] while the view of a molecule as a mere set of atoms formed the basis for the theory of refraction.[14] More recently, considering a molecule or a material as a combination of "building blocks" – bonds or groups of atoms – has led to the development of predictive group additivity methods,[15-18] scaling relationships,[19-26] $O(N)$ methods for accelerating quantum calculations,[27-29] and chemical bonding analysis tools.[30-36] The localized view on chemical bonding has driven the creation of low-cost methods for computing correlation energy[37, 38] and continues to inspire highly successful machine learned interatomic potentials[39-43] and data-driven density functional theory (DFT) Hamiltonians.[44-46]

The successful applications of local bonding concepts across various fields hint at the existence of a universal theory of localized electronic states that could reconcile both local and non-local electronic phenomena and potentially reduce the cost of quantum calculations. Adams (1961),[47] Gilbert (1964),[48] and Anderson (1968)[49] independently proposed eigenvalue problems for localized orbitals. Adams-Gilbert and Anderson equations were tested for band structure calculations by Kunz[50, 51] and Bullett,[52-54] respectively, as well as by others, with mixed success.[55-57] The modified Adams-Gilbert approach[58, 59] formed the basis for the Head-Gordon's ALMO-EDA energy decomposition method,[60-67] which clarified the physics of a chemical bond,[68] while Anderson's "chemical pseudopotential" method gave rise to all past and current empirical bond order potentials and reactive force fields[69-74] through the pioneering work of Abell (1985).[75] A remarkable connection between the localized-state eigenproblem and the orthogonal tight binding theory, discovered by Anderson,[76] suggests that the universal theory of localized states, if it exists, likely resembles the tight binding formalism.

Empirical tight binding (TB) methods approximate the electronic structure problem by employing minimal atomic basis sets, neglecting multicenter integrals, and replacing difficult-to-compute terms with simpler, parameterized expressions. As a result, their cost can be up to three orders of magnitude lower than that of standard DFT.[77] For molecules, Burrau (1927),[78] Condon (1927),[79] Lennard-Jones (1929),[80] and Hückel (1931)[81] introduced the linear combination of atomic states (LCAO) TB method to describe molecular electronic states, first elucidated by Hund (1927)[82-85] and Mulliken (1928).[86, 87] The resulting molecular orbital (MO) theory was further advanced by Coulson and Longuet-Higgins,[88-93] Dewar,[94] Mulliken,[95-97] and Löwdin.[98] Notable empirical resonance integral expressions for molecules were proposed by Lennard-Jones (1937),[99] Mulliken (1948),[100] and Wolfsberg and Helmholtz (1952).[101] For periodic systems, empirical TB methods were pioneered by Bloch (1929)[102] and Slater and Koster (1954)[103] for energy band interpolations, culminating with the systematic work of Harrison.[104] Inverse-power resonance integrals used in periodic TB were first proposed by Phillips (1970).[105] Roothaan (1948),[106, 107] Lennard-Jones (1949),[108] and Hall (1950)[109, 110] introduced the non-empirical LCAO method, which was simplified by Pariser, Parr,[111, 112] and Pople[113] in the form of the self-consistent empirical MO theory (1953). Fletcher and Wohlfarth were likely the first to perform the non-empirical LCAO TB calculation for band structure (1951).[114] Studies of total energy differences and geometries at low cost became possible with the advent of Hoffmann's extended Hückel (eH) method (1963),[115] improved by A.B. Anderson,[116, 117] and the introduction of Pople's CNDO, NDDO, and INDO approximations to Hartree-Fock-Roothaan equations (1965).[118-121] Pople's semiempirical methods were further developed by Dewar, Thiel, and Stewart.[122-129] Total energy TB calculations for periodic solids were pioneered by Messmer and Watkins (1970)[130] using the eH method at the Γ-point, by Bullett[53] and Anderson[131] using similar methods with added repulsive corrections (1975), and by Chadi (1978)[132] using repulsive terms and band integration.

Several TB theories have been systematically developed by simplifying the Kohn-Sham DFT (KS-DFT)[133] equations. Andersen and Jepsen transformed the LMTO-ASA[134] approximation of KS-DFT into the TB form (1984).[135] However, due to the use of space-filling spheres as a basis set in the LMTO-ASA formalism, the method was largely restricted to close-packed structures.[136] Sutton et al. (1988)[137] and Foulkes and Haydock (1989)[138] derived TB equations from the Harris-Foulkes approximation[139, 140] to KS-DFT. Sankey and Niklewski similarly developed a practical *ab initio* TB scheme (1989),[141] which has found widespread use[142, 143] and has undergone several stages of development.[144-150] Like all related TB methods, it is not fully self-consistent due to reliance on the Harris-Foulkes approximation. Additionally, it was reported to be 1-2 orders of magnitude slower[151] than empirical TB due to the retention of three-center integrals.

At present, state-of-the-art empirical TB methods include two-center and non-orthogonal DFTB by Seifert, Elstner, and Frauenheim (1995),[152-169] two-center and non-orthogonal GFN-xTB by Grimme and Bannwarth (2017),[170-173] and three-center, orthogonal $OM_x$ by Thiel (1993).[174-180] Accuracy of recent $OM_x$ and DFTB versions rivals that of the DFT generalized gradient approximation,[181, 182] which is achieved by extensive parameterization against large training datasets.

Recently, the author reported a discovery of a set of self-consistent, orthogonal tight binding-type equations that describe the energetics of $H_x$ model systems more accurately than conventional empirical tight binding techniques *using analytical two-center resonance integrals and no parameters*.[183] This revelation challenges the prevailing notion that tight binding methods in their common two-center form are inherently empirical. The non-empirical nature of the reported model suggests the possibility of a new *ab initio* theory founded on principles distinct from the standard independent electron ansatz.

In the present work, a formally exact localized-orbital/tight-binding framework is established that fundamentally explains the success of the earlier non-empirical results in Ref. 183, hereafter referred to as Paper I. A new DFT reference state – the independent atom ansatz – is introduced that allows for the representation of electron density in terms of non-interacting, perturbed atomic orbitals. The ansatz enables natural classification of exchange-correlation effects into intra-atomic and inter-atomic contributions and facilitates partial cancellation of inter-atomic repulsive electron-electron and attractive electron-nuclear interactions. This cancellation validates the use of the small-overlap, weak interaction limit for obtaining mathematical forms of inter-atomic terms. The derived self-consistent equations bear resemblance to the theory of localized orbitals proposed by P.W. Anderson,[49, 184] and simultaneously to the charge equilibration (QEq) method by Rappe and Goddard.[185] It is notable that Hückel theory, Sanderson's electronegativity equalization principle,[186] and atomic Aufbau principle emerge naturally in the framework presented. The method provides mechanisms for removal of self-interaction and static correlation errors and incorporates atomic charge and energy decomposition analyzes at no cost.

The paper is organized as follows. In Section II, reference systems for modeling many-body effects are discussed, and the independent atom ansatz is introduced. In Section III, its total energy expression is derived in the most general form by applying the two-dimensional adiabatic connection formalism to the second-quantized Schrödinger equation. In Section IV, self-consistent equations are introduced using the method of constrained optimization. In Section V, an inter-atomic exchange-correlation functional and its corresponding potential are derived using asymptotic correspondence among electronic structure theories. In Section VI, self-consistency and formal exactness of derived equations are discussed. In Section VII, a mathematical form of the resonance integral is obtained. In Section VIII, a comparison is made with existing tight binding models, and self-interaction and static correlation errors are discussed. In Section IX, model re-interpretations motivated by physical constraints are analyzed. Finally, conclusions are presented in Section X.

# II. REFERENCE SYSTEMS FOR MODELING MANY-BODY EFFECTS

## A. Comparison of references in statistical mechanics and quantum mechanics

Practical methods for predicting the observable properties of many-body systems, both classical and quantum mechanical, commonly rely on computationally tractable reference systems. In these methods, the mean value of an observable $\langle f\rangle$ is expressed as the sum of its value in the reference system $\langle f\rangle_{ref}$ and a correction term $\langle f\rangle_{corr}$:

$$\langle f\rangle = \langle f\rangle_{ref} + \langle f\rangle_{corr}. \quad (1)$$

This property representation finds wide application in classical thermodynamics[187] and statistical mechanical theories of liquids and liquid mixtures,[188-190] as well as in nearly all practical *ab initio* quantum chemistry methods.[1133] In Kohn-Sham density functional theory (KS-DFT), $f \equiv E[\rho]$ is the sum of the energy of an independent electron gas and the correction term – the exchange-correlation (XC) functional $E_{xc}[\rho]$.

In theories of liquids, it is acknowledged that the smallness and simplicity of $\langle f\rangle_{corr}$ in eq. (1) depend on the similarity between the reference system and the system being modeled. For a dense monoatomic liquid as the system and an ideal gas as the reference, $\langle f\rangle_{corr}$ is highly complex and forms divergent series.[191] In contrast, for a liquid-like Weeks-Chandler-Andersen reference state, $\langle f\rangle_{corr}$ takes a simple, first-order perturbative form, enabling accurate predictions of free energies and pressures for simple liquids.[188]

In contrast to statistical mechanics, in quantum mechanics, the consideration of similarity between a system and a reference to simplify $\langle f\rangle_{corr}$ or $E_{xc}[\rho]$ has received limited attention, primarily discussed in relation to multireference methods,[192] to the best of my knowledge. The widely used KS-DFT method employs the same "ideal electron gas" reference state to describe systems with qualitatively different electron density profiles, such as metals with delocalized electrons and molecules or insulators with electron localization. The non-interacting electron reference is used despite the fact that electron densities span more than two orders of magnitude – the typical range in classical fluids – within a single atom alone.[193]

## B. Limitations of the independent electron ansatz in density functional theory

It can be argued that the challenges[194] associated with designing a universal XC functional $E_{xc}[\rho]$ in DFT can be attributed, at least in part, to the use of the non-optimal independent electron reference state. Employing the adiabatic connection formalism[195, 196] and defining $H_\lambda = T + V_{ref} + \lambda V_{pert} + V_\lambda$, where $V_{ref}$, $V_{pert}$, and $V_\lambda$ are the reference, perturbation, and constraining potentials, it is evident that in the expansion of $H_\lambda$ eigenfunctions $|\Psi_\lambda\rangle$ in the representation of $H_{\lambda=0}$ eigenstates, many coefficients would be non-zero and non-analytical if $V_{pert}$ is large. Consequently, the evaluation of the integral in $E_{xc}[\rho] = \int_{\lambda=0}^{\lambda=1}\langle\Psi_\lambda|V_{pert}|\Psi_\lambda\rangle d\lambda - J[\rho]$, where $J$ is the standard Hartree integral, would be very challenging. This is the case in the KS theory, where $V_{pert}$ corresponds to strong electron-electron repulsion, corroborating the observation that finding a simple and accurate $E_{xc}[\rho]$ expression is a highly non-trivial task.[194]

If a reference system is chosen to closely resemble the system under analysis, the smallness of $V_{pert}$ shall enable construction of a rapidly convergent perturbative expansion of $|\Psi_\lambda\rangle$ in $\lambda$, making it possible to evaluate the $E_{xc}[\rho]$ integral analytically, at least in principle.

### C. Independent atom ansatz

In this work, a new DFT reference state is proposed to reduce $V_{pert}$ and thereby simplify DFT calculations. The approach relies on the straightforward concept that molecules are composed of atoms and defines non-interacting, perturbed atomic orbitals as a reference. These orbitals are constructed to reproduce the exact electron density. The corresponding ansatz is referred to as the *independent atom DFT ansatz*. The ansatz requires that $V_{pert}$ contains both repulsive electron-electron and attractive nuclear-electron *inter-atomic* interactions. Since they have opposite signs, they partially cancel each other, reducing the net $V_{pert}$ effect and thus simplifying the $E_{xc}$ evaluation.

The formally exact general forms of equations obtained from the independent atom ansatz in Sections III-VI resemble tight-binding theories and are collectively referred to as *non-empirical tight binding theory*, or *NTB*, in this work. In the next section, the NTB total energy expression is derived in the most general form from the second-quantized Schrödinger equation using the two-dimensional adiabatic connection formalism.

## III. TOTAL ENERGY IN THE INDEPENDENT ATOM ANSATZ

### A. Second-quantized Hamiltonian for overlapping atomic states

The exact nonrelativistic ground-state energy $E$ and the wave function $|\Psi\rangle$ are the solutions to the Schrödinger equation:

$$H|\Psi\rangle = E|\Psi\rangle, \quad (2)$$

where the Hamiltonian $H$ of a system of $M$ nuclei at $\mathbf{R}_a$ positions and $N$ electrons has the following second-quantized form:[1]

$$H = \sum_{ij}\langle i|t|j\rangle a_i^\dagger a_j + \sum_{ija}\langle i|v_a|j\rangle a_i^\dagger a_j + \frac{1}{2}\sum_{ijkl}\langle ij|kl\rangle a_i^\dagger a_j^\dagger a_l a_k + E_{NN}, \quad (3)$$

where $a_i^\dagger$ and $a_i$ are creation and annihilation operators, respectively; $t = -\frac{1}{2}\nabla^2$; $|i\rangle$ states are assumed orthogonal; $v_a$ is the electrostatic potential due to nucleus $a$ with charge $Z_a$, $\langle ij|kl\rangle$ are the standard two-electron integrals between spin orbitals,[1] and $E_{NN}$ is the standard internuclear interaction energy.

In the following, we relax the orthogonality requirement of the $|i\rangle$ states and choose them to be localized on atoms. We define elements of the overlap matrix as follows:

$$S_{ab} = \langle a|b\rangle - \delta_{ab}, \quad (4)$$

so that $S_{aa} = 0$. Then, eq. (3) is modified as follows:[197]

$$\begin{aligned} H = &\sum_{abcd}(\mathbf{I}+\mathbf{S})^{-1}_{ca}\langle a|t|b\rangle(\mathbf{I}+\mathbf{S})^{-1}_{bd}a_c^\dagger a_d \\ &+\sum_{abcde}(\mathbf{I}+\mathbf{S})^{-1}_{ca}\langle a|v_e|b\rangle(\mathbf{I}+\mathbf{S})^{-1}_{bd}a_c^\dagger a_d \\ &+\frac{1}{2}\sum_{\substack{abcd\\efgk}} \begin{matrix}(\mathbf{I}+\mathbf{S})^{-1}_{ea}(\mathbf{I}+\mathbf{S})^{-1}_{fb}\langle ab|cd\rangle \\ \times(\mathbf{I}+\mathbf{S})^{-1}_{gc}(\mathbf{I}+\mathbf{S})^{-1}_{kd}a_e^\dagger a_f^\dagger a_k a_g\end{matrix} \\ &+E_{NN}, \end{aligned} \quad (5)$$

where $\mathbf{I}$ and $\mathbf{S}$ are the identity and overlap matrices, respectively.

In eq. (5), three groups of terms correspond to the kinetic energy $T$, the nuclear-electron potential energy operator $V_{Ne}$, and the electron-electron potential energy operator $V_{ee}$, respectively. Next, we employ the Löwdin expansion[98] of the $(\mathbf{I}+\mathbf{S})^{-1}$ matrix elements:

$$(\mathbf{I}+\mathbf{S})^{-1}_{ab} = \delta_{ab} - S_{ab} + \sum_c S_{ac}S_{cb} - \sum_{cd} S_{ac}S_{cd}S_{db} + - \cdots. \quad (6)$$

After substitution into eq. (5), the resulting terms corresponding to $V_{Ne}$ and $V_{ee}$ can be classified based on whether the orbital indices they involve belong to only one atom or more than one atom. Under this classification, the potential energy terms belonging to one atom at a time are grouped into $M$ intra-atomic nuclear-electron ($V_{Ne}^a$) and electron-electron ($V_{ee}^a$) groups. The remaining terms corresponding to more than one atom are assigned to the inter-atomic groups $V_{Ne}^{\circ\circ}$ and $V_{ee}^{\circ\circ}$, respectively. After this grouping procedure, Eq. (5) becomes

$$H = -\sum_i \frac{1}{2}\nabla_i^2 + \sum_a V_{Ne}^a$$
$$+\sum_a V_{ee}^a + V_{Ne}^{\circ\!-\!\circ} + V_{ee}^{\circ\!-\!\circ} + E_{NN}. \quad (7)$$

In eq. (7), the second-quantized kinetic energy operator was replaced with its non-quantized form. This Hamiltonian form provides a convenient starting point for implementing the adiabatic connection method that leads to the NTB theory, as described next.

### B. Adiabatic connection for the second-quantized Hamiltonian

A new Hamiltonian of the following form is introduced based on eq. (7):

$$H_{\lambda_1\lambda_2} = \sum_i -\frac{1}{2}\nabla_i^2 + v_{\lambda_1\lambda_2}$$
$$+\lambda_1 \sum_a V_{ee}^a + \lambda_2 (V_{Ne}^{\circ\!-\!\circ} + V_{ee}^{\circ\!-\!\circ} + E_{NN}). \quad (8)$$

Here, $\lambda_1$ and $\lambda_2$ are the coupling constants linking interacting and non-interacting systems, and $v_{\lambda_1\lambda_2}$ is the constraining potential that ensures the electron density equals the exact density for any sets of $\lambda$ values. Its two limiting cases are:

$$v_{11} = \sum_a V_{Ne}^a\,,$$
$$v_{00} = \sum_{ak} v_{HA}^{ak}\,, \quad (9)$$

where $v_{HA}^{ak}$ is the set of effective, non-local, orbital-dependent potentials determining the shapes of non-interacting orbitals $|\varphi_{ak}\rangle$ (*vide infra*) in the absence of $V_{ee+Ne}^{\circ\!-\!\circ}$ and $V_{ee}^a$ interactions.

Interacting and non-interacting systems can be connected using any convenient path in the $\{\lambda\}$ space. Inspired by the evolution of atoms and molecules during the early stages of the Universe (one-electron atoms → many-electron atoms → complex molecules), a path is constructed in which the intra-atomic interactions $V_{ee}^a$ are enabled first through an increase in $\lambda_1$ from 0 to 1, followed by the activation of $(V_{Ne}^{\circ\!-\!\circ} + V_{ee}^{\circ\!-\!\circ} + E_{NN})$ by $\lambda_2$. $E_{NN}$ is included to mimic the physical process of molecule formation from atoms and does not affect the wave function's shape.

The total energy of the interacting system then becomes:

$$E = E_{11} = E_{00} + \int_{\lambda_1=0}^{\lambda_1=1} \left(\frac{\partial E}{\partial \lambda_1}\right)_{\lambda_1 0} d\lambda_1$$
$$+\int_{\lambda_2=0}^{\lambda_2=1} \left(\frac{\partial E}{\partial \lambda_2}\right)_{1\lambda_2} d\lambda_2\,, \quad (10)$$

where the subscripts denote $\lambda_1$ and $\lambda_2$ in this order. The $E_{00}$ term in eq. (10) is the energy of the non-interacting system in the independent atom ansatz with the electron density constrained to reproduce the exact density.

### C. Reference system of non-interacting atoms

In the non-interacting reference system, the electron density equals the sum of perturbed densities of atoms $\rho_a$:

$$\rho(\mathbf{r}) = \sum_a \rho_a(\mathbf{r})\,. \quad (11)$$

The atomic densities $\rho_a$, in turn, are the superpositions of perturbed atomic orbital densities $\rho_{ak}$, weighted by their occupancies $q_{ak}$:

$$\rho_a(\mathbf{r}) = \sum_k q_{ak}\rho_{ak}(\mathbf{r})\,, \quad (12)$$

where $k$ is the quantum state index in atom $a$. $\rho_{ak}$ is the squared modulus of a perturbed atomic orbital $\varphi_{ak}$:

$$\rho_{ak}(\mathbf{r}) = |\varphi_{ak}(\mathbf{r})|^2. \quad (13)$$

The $\rho_{ak}$ quantities satisfy the normalization constraint:

$$q_{ak}\int \rho_{ak}(\mathbf{r})d\mathbf{r} - q_{ak} = 0\ \forall ak, \quad (14)$$

where we multiplied both terms by $q_{ak}$ to facilitate the subsequent analysis in Section IV. The $q_{ak}$ values obey the charge conservation condition:

$$\sum_{ak} q_{ak} - N = 0 \quad (15)$$

and the Pauli exclusion principle:

$$q_{ak} - 2 \leq 0 \; \forall ak \tag{16}$$

and

$$-q_{ak} \leq 0 \; \forall ak. \tag{17}$$

The perturbed atomic states $|\varphi_{ak}\rangle$ are the solutions to the following eigenvalue problem:

$$\begin{aligned} H_{HA}^{ak}|\varphi_{ak}\rangle &= \varepsilon_{ak}|\varphi_{ak}\rangle, \\ H_{HA}^{ak} &= -\frac{1}{2}\nabla^2 + v_{HA}^{ak}, \end{aligned} \tag{18}$$

where $v_{HA}^{ak}$ is the constraining potential, same as in eq. (9), that ensures that $\rho$ in eq. (11) equals the exact density of the system. The total energy of the reference system containing non-interacting perturbed atomic orbitals is then

$$E_{00} = \sum_{ak} q_{ak} \left\langle \varphi_{ak} \middle| -\frac{1}{2}\nabla^2 + v_{HA}^{ak} \middle| \varphi_{ak} \right\rangle \tag{19}$$

### D. Total energy functional

After substituting eq. (19) into eq. (10), employing the Hellmann-Feynman theorem[198, 199] to expand the partial derivatives, and integrating over $\lambda_1$ and $\lambda_2$, the following NTB total energy functional is obtained:

$$\begin{aligned} E[\rho] = \sum_{ak} q_{ak} \left\langle \varphi_{ak} \middle| -\frac{1}{2}\nabla^2 \middle| \varphi_{ak} \right\rangle + E_{es}[\rho] \\ + \sum_{a} E_{xc}^{a}[\rho_a] + E_{xc}^{\infty}[\rho]. \end{aligned} \tag{20}$$

Here, new intra-atomic $E_{xc}^{a}$ and inter-atomic $E_{xc}^{\infty}$ XC functionals in the independent atom ansatz – so-called *IA-XC functionals* – are defined as

$$\begin{aligned} E_{xc}^{a}[\rho] &= \int_{\lambda_1=0}^{\lambda_1=1} \langle \Psi_{\lambda_1 0} | V_{ee}^{a} | \Psi_{\lambda_1 0} \rangle d\lambda_1 - J[\rho_a], \\ E_{xc}^{\infty}[\rho] &= \int_{\lambda_2=0}^{\lambda_2=1} \left\langle \Psi_{1\lambda_2} \middle| \begin{matrix} V_{Ne}^{\infty} + V_{ee}^{\infty} \\ + E_{NN} \end{matrix} \middle| \Psi_{1\lambda_2} \right\rangle d\lambda_2 \\ &\quad - E_{es}^{\infty}[\rho], \end{aligned} \tag{21}$$

where $E_{es}^{\infty}$ is the inter-atomic electrostatic energy:

$$\begin{aligned} E_{es}^{\infty}[\rho] = \sum_{ak, b\neq a} q_{ak} \int \rho_{ak}(\mathbf{r})\, v_b(\mathbf{r}) d\mathbf{r} \\ + \frac{1}{2} \sum_{ak, be, b\neq a} q_{be} q_{ak} \int \frac{\rho_{be}(\mathbf{r})\rho_{ak}(\mathbf{r}')}{|\mathbf{r}-\mathbf{r}'|} d\mathbf{r} d\mathbf{r}' \\ + E_{NN}. \end{aligned} \tag{22}$$

The form of $E_{xc}^{\infty}[\rho]$ is obtained in Sections V-VII. The total electrostatic energy $E_{es}$ in eq. (20) is defined as

$$\begin{aligned} E_{es}[\rho] = E_{es}^{\infty}[\rho] \\ + \sum_{a} \left( \int \rho_a(\mathbf{r})\, v_a(\mathbf{r}) d\mathbf{r} + J[\rho_a] \right). \end{aligned} \tag{23}$$

The above expressions involving atom-localized $|\varphi_{ak}\rangle$ states resemble those containing the KS $|\psi_i\rangle$ states in KS-DFT. Analogous to $|\psi_i\rangle$ that are often referred to as quasiparticles in the solid state physics literature,[200] I refer to the $|\varphi_{ak}\rangle$ states as quasiparticles and call them *atomions* in this work. The name is derived from "atom" and "fermion". I refer to $\varepsilon_{ak}$ as *atomion energies*, $\rho_{ak}$ as *atomion densities*, and $q_{ak}$ as *atomion occupancies.*

## IV. SELF-CONSISTENT EQUATIONS

### A. Constrained optimization problem

The NTB ground-state total energy functional in eq. (20) is formally exact and thus satisfies the 1st and 2nd Hohenberg-Kohn theorems[201] – it is a functional of the electron density that minimizes it. Since $\rho = \sum_{ak} q_{ak}\rho_{ak}$ (cf. eq. (11) and (12)), $E$ is also minimized with respect to variations in $\rho_{ak}$ and $q_{ak}$ subject to eq. (14)-(17) constraints. The ground state can then be obtained by the unconstrained minimization of the generalized Lagrangian functional:

$$\begin{aligned} \mathcal{L} = E[\{q_{ak}\}, \{\rho_{ak}\}] \\ - \sum_{ak} \varepsilon_{ak} \left[ q_{ak} \int \rho_{ak}(\mathbf{r}) d\mathbf{r} - q_{ak} \right] \\ + \lambda \left( \sum_{ak} q_{ak} - N \right) + \sum_{ak} \eta_{ak}(q_{ak} - 2) \\ + \sum_{ak} \nu_{ak}(-q_{ak}), \end{aligned} \tag{24}$$

where $\varepsilon_{ak}$ and $\lambda$ are the Lagrange multipliers, and $\eta_{ak}$ and $\nu_{ak}$ are the Karush-Kuhn-Tucker (KKT) multipliers.

At the minimum,

$$\frac{\delta \mathcal{L}}{\delta \rho_{ak}(\mathbf{r})} = 0,$$
$$\frac{\partial \mathcal{L}}{\partial q_{ak}} = 0. \tag{25}$$

Substitution of eq. (24) into eq. (25) and division of the first equation by $q_{ak}$ results in

$$\frac{1}{q_{ak}} \frac{\delta E}{\delta \rho_{ak}(\mathbf{r})} = \varepsilon_{ak} \tag{26}$$

and

$$\frac{\partial E}{\partial q_{ak}} = -\lambda - \eta_{ak} + \nu_{ak}. \tag{27}$$

Solving eq. (26) and (27) simultaneously shall yield $\{\rho_{ak}\}$ and $\{q_{ak}\}$ in the ground state of the system.

## B. Atomion eigenvalue problem

In Section IIIC, eq. (18), I introduced the one-electron eigenvalue problem for atomions $|\varphi_{ak}\rangle$ that contained the unknown constraining potential $v_{HA}^{ak}$. In order to determine the form of $v_{HA}^{ak}$, we substitute eq. (20) into (26), take the derivative with respect to $\delta\rho_{ak} = \delta\varphi_{ak}^{*}\varphi_{ak}$, divide the result by $q_{ak}$, and right-multiply both sides by $|\varphi_{ak}\rangle$ to obtain

$$\left[-\frac{1}{2}\nabla^2 + \sum_{b} v_{es}^{b} + \mu_{xc}^{ak} + \mu_{xc}^{\infty,ak}\right]|\varphi_{ak}\rangle = \varepsilon_{ak}|\varphi_{ak}\rangle, \tag{28}$$

where $v_{es}^{b} = v_b(\mathbf{r}) + \phi_b(\mathbf{r})$ is the electrostatic potential due to atom $b$ with the atomic Hartree potential $\phi_b(\mathbf{r})$. $\mu_{xc}^{ak}$ and $\mu_{xc}^{\infty,ak}$ are intra-atomic and inter-atomic XC potentials, respectively, acting on $|\varphi_{ak}\rangle$. They are defined as

$$\mu_{xc}^{ak} = \frac{1}{q_{ak}} \frac{\delta E_{xc}^{a}[\rho_a]}{\delta \rho_{ak}},$$
$$\mu_{xc}^{\infty,ak} = \frac{1}{q_{ak}} \frac{\delta E_{xc}^{\infty}[\rho]}{\delta \rho_{ak}}. \tag{29}$$

A comparison of eq. (18) and (28) reveals that

$$v_{HA}^{ak} = \sum_{b} v_{es}^{b} + \mu_{xc}^{ak} + \mu_{xc}^{\infty,ak}. \tag{30}$$

## C. Atomion equalization principle

Having obtained the general form of the eigenvalue problem that yields atomions $|\varphi_{ak}\rangle$ as its solutions (eq. (28)), it remains to develop equations for computing $\{q_{ak}\}$. After substituting eq. (20) into the left-hand side of eq. (27) and taking the derivative, it follows that

$$\frac{\partial E}{\partial q_{ak}} = \left\langle \varphi_{ak} \middle| -\frac{1}{2}\nabla^2 + \textstyle\sum_b v_{es}^{b} \middle| \varphi_{ak} \right\rangle + \frac{\partial E_{xc}^{a}}{\partial q_{ak}} + \frac{\partial E_{xc}^{\infty}}{\partial q_{ak}}. \tag{31}$$

Since $\partial E_{xc}[\rho]/\partial q_{ak} = \langle \varphi_{ak}|\mu_{xc}|\varphi_{ak}\rangle$ (see Section S1), it follows from eq. (31), eq. (27), and the integral form of eq. (28) that

$$\frac{\partial E}{\partial q_{ak}} = \varepsilon_{ak} \tag{32}$$

and

$$\varepsilon_{ak} = -\lambda - \eta_{ak} + \nu_{ak}. \tag{33}$$

When $q_{ak} = 2$, the eq. (16) constraint is active and $\eta_{ak} > 0$, according to the dual feasibility KKT condition.[202] Similarly, $\nu_{ak} > 0$ when $q_{ak} = 0$. Therefore, if $\varepsilon_{a1}$ is fully occupied, $\varepsilon_{a2}$ is partially occupied, and $\varepsilon_{a3}$ is empty, we have $\varepsilon_{a1} < \varepsilon_{a2} < \varepsilon_{a3}$ – the atomion Aufbau principle naturally follows. Evidently, the Aufbau principle of KS-DFT and the Hartree-Fock (HF) theory can be derived in a similar manner if one replaces atomions $\varphi_{ak}$ with KS or HF states $\psi_{ak}$ and use either $E_{xc}$ or $E_x$ in place of the IA-XC functionals in eq. (20). After completion of this work, it came to my attention that Giesbertz and Baerends have already derived that standard Aufbau principle from KKT conditions in a similar way back in 2010.[203]

Atoms that participate in chemical bonding host $0 < q_{ak} < 2$ electrons on their highest occupied atomions with energies $\varepsilon_{ak}$. The corresponding partially occupied atomions are referred to as *valence atomions*. Since the KKT inequality constraints are inactive for this range of $q_{ak}$ values, we have $\eta_{ak} = 0$, $\nu_{ak} = 0$, and thus $\varepsilon_{ak} = -\lambda$, according to eq. (33). It then follows that all partially occupied atomions in a molecule are degenerate:

$$\varepsilon_{ak} = \varepsilon_{am} = \varepsilon_{be} = \cdots = \varepsilon,$$
$$if\ q_{ak}, q_{am}, q_{be}, \ldots \in (0,2) \tag{34}$$

I refer to this remarkable corollary as *the atomion equalization principle.*

Occupancies of the valence atomions (denoted by $q_{be}^{val}$) can be found by solving the matrix equation. It is obtained by representing $\varepsilon_{ak}$ in the following form:

$$\varepsilon_{ak} = \varepsilon_{ak}^{0}(\{q_{be}^{val}\}) + \sum_{be} q_{be}^{val} \mathcal{F}_{beak}\,, \tag{35}$$

where

$$\varepsilon_{ak}^{0} = \left\langle \varphi_{ak} \middle| -\frac{1}{2}\nabla^2 + \textstyle\sum_b v_b \middle| \varphi_{ak} \right\rangle + \frac{\partial E_{xc}^{a}}{\partial q_{ak}} + \frac{\partial E_{xc}^{\infty}}{\partial q_{ak}} \tag{36}$$

and

$$\mathcal{F}_{beak} = \int \frac{\rho_{be}(\mathbf{r})\rho_{ak}(\mathbf{r}')}{|\mathbf{r}-\mathbf{r}'|} d\mathbf{r}d\mathbf{r}'\,. \tag{37}$$

After taking into account eq. (34) and the fact that $\sum_{ak} q_{ak}^{val} = N_v$, where $N_v$ is the number of valence electrons in the system, the following matrix equation is obtained:

$$\mathbf{Kq} = \mathbf{b}, \tag{38}$$

where

$$\begin{gathered} K_{ij} = 1 \text{ if } i = 1, \\ K_{ij} = \mathcal{F}_{ij} - \mathcal{F}_{1j} \text{ if } i > 1 \\ b_1 = N_v, \\ b_i = \varepsilon_1^0 - \varepsilon_i^0, i > 1, \\ q_i = q_{ak}^{val}. \end{gathered} \tag{39}$$

Since $\varepsilon_{ak}^{0}$ is in general charge-dependent, the matrix equation must be solved self-consistently. Eq. (34), (35), and (38) strongly resemble Rappe-Goddard's QEq scheme for computing atomic charges.[185] The key difference is that the current equations are formally exact, provided that IA-XC functionals and their corresponding potentials are known, and charge optimization is coupled with relaxation of atomions through eq. (28).

The common atomion energy value $\varepsilon$ is identical to the electronic chemical potential $\mu$ of the system – this follows from the fractional charge argument if it is extended from Cohen et al.[204] to the independent atom ansatz (see Section S2). Since $-\mu$ is referred to as electronegativity, the quantities such as $-\varepsilon_{ak}^{val}$ correspond to atomic electronegativities that equalize in bonded systems, according to eq. (34), providing a physical justification for the empirical Sanderson's electronegativity equalization principle.[186] Curiously, the quantity $-\varepsilon_{ak}^{val}$ in *sp* atoms is equivalent to the Allen electronegativity,[205] if the s and p shells are partially occupied and thus degenerate, as in *sp*[3]-hybridized carbon-containing molecules. In metals with partially occupied d-states, $\varepsilon_{ak}^{val}$ should correlate with the atom-projected d-band center[206] – a highly important descriptor for the bonding strength between chemisorbed species and metallic surfaces.

Parr and co-workers studied a connection between atomic electronegativities and chemical potentials by either considering systems of weakly interacting atoms or using approximate bond charge models.[207, 208] The formalism presented herein can be regarded as a generalization of such earlier developments to strongly interacting atoms, providing a theoretical foundation for the observed electronegativity equalization in actual molecules.

The utility of equations obtained so far hinges on the availability of $E_{xc}^{\infty}$ and $\mu_{xc}^{\infty,ak}$ functional forms that are derived next.

# V. INTER-ATOMIC EXCHANGE AND CORRELATION FUNCTIONAL

## A. *Translatio ex infinitum* method

In this section, I introduce the method of *translatio ex infinitum* ("translation from infinity") to obtain the functional form of the inter-atomic IA-XC potential $\mu_{xc}^{\infty,ak}$ and the corresponding IA-XC functional $E_{xc}^{\infty}$. In this approach, mathematical forms of the inter-atomic NTB terms are obtained in the $R \to \infty$ limit using the *asymptotic correspondence principle* – unification of inter-atomic energy expressions in KS-DFT, HF/CI, NTB, and valence bond (VB) theories in the limit of small inter-atomic overlaps ($S \to 0$). In Section VB, the transferability of the asymptotic expressions to typical $R$ and $S$ values in chemical bonds is justified. In Section VC, the asymptotic correspondence principle is formally derived and discussed.

The *translatio ex infinitum* technique was inspired by the correspondence principle between quantum and classical mechanics in the small wavelength limit that has been historically employed to constrain the quantum mechanical formalism.[209] The NTB expressions obtained using the *translatio ex infinitum* technique make up an internally consistent theory (see Sections VI-IX) and have been demonstrated for simple model systems to be numerically accurate (see Paper I[183]).

## B. Why is the small overlap limit appropriate?

In Section IIID, the inter-atomic IA-XC functional was defined as

$$E_{xc}^{\infty\infty}[\rho] = \int_{\lambda_2=0}^{\lambda_2=1} \langle \Psi_{1\lambda_2} | V_{Ne}^{\infty\infty} + V_{ee}^{\infty\infty} + E_{NN} | \Psi_{1\lambda_2} \rangle \times d\lambda_2 - E_{es}^{\infty\infty}[\rho], \quad (40)$$

where the coupling constant $\lambda_2$ turns on inter-atomic interactions and effectively mimics the molecule formation from free atoms. The $\lambda_2$-dependence of $|\Psi_{1\lambda_2}\rangle$ in eq. (40) emerges under the action of the combined $ee + Ne$ inter-atomic potential $V_{Ne}^{\infty\infty} + V_{ee}^{\infty\infty}$ (cf. eq. (8)). In comparison with KS-DFT, where the $V_{ee}$ potential perturbation is strong, the $V_{Ne}^{\infty\infty} + V_{ee}^{\infty\infty}$ perturbation is considerably weaker as the $ee$ and $Ne$ interactions have opposite signs. For example, in the $H_2$ molecule at the equilibrium bond length, the electrostatic energy of electron-1 located near nucleus-1 in the combined field of electron-2 and nucleus-2 equals -2.90 eV, when computed with the methods of Paper I.[183] This is considerably lower in magnitude than the electron-1 energy in the field of electron-2 alone (+13.70 eV), the total potential energy experienced by electron-1 (-29.91 eV), or the energy difference between unperturbed 1s and 2s H atom states (10.20 eV).

Due to the smallness of $V_{Ne}^{\infty\infty} + V_{ee}^{\infty\infty}$, the dependence of $|\Psi_{1\lambda_2}\rangle$ on $\lambda_2$ can be expected to be quite weak. The unknown general form of $|\Psi_{1\lambda_2}\rangle$ can therefore be represented as a Taylor series in $\lambda_2$ up to the linear term, with $|\Psi_{10}\rangle$ being the leading term that contains products of atomic orbitals. Since $\lambda_2 = 0$ corresponds to inter-atomic distances $R \to \infty$ and $S \to 0$, it is thus natural to use the $S \to 0$ limit to obtain $\mu_{xc}^{\infty\infty,ak}$ and $E_{xc}^{\infty\infty}[\rho]$ expressions. The inclusion of terms linear in $\lambda_2$ will then be equivalent to incorporating the $O(S)$ terms into $E_{xc}^{\infty\infty}$ and $\mu_{xc}^{\infty\infty,ak}$. Due to the partial cancellation of the $V_{Ne}^{\infty\infty}$ and $V_{ee}^{\infty\infty}$ effects, the resulting expressions are expected to be accurate even when the overlap $S$ is no longer small, corroborating the *translatio ex infinitum* method introduced in Section VA.

## C. Asymptotic correspondence of electronic structure theories

*Asymptotic correspondence to $O(S^0)$*. To illustrate the asymptotic correspondence of the formally exact electronic structure theories (NTB, HF/CI, KS-DFT, and VB) to the lowest order in the differential overlap $\varphi_a(\mathbf{r})\varphi_b(\mathbf{r})$, we note that at $R \gg 0$, a molecule is a superposition of weakly interacting, nearly free atoms. Assuming one orbital per atom for simplicity, every energy term can be expressed as $E = E[\{\varphi_a\}, \{\varphi_a\varphi_b\}]$. Taylor expanding $E$ with respect to small $\varphi_a\varphi_b$ up to linear terms yields $E = \sum_a E[\varphi_a] + \sum_{a<b} E^{\leftrightarrow}[\varphi_a\varphi_b]$, where the second term reflects the summation of pairwise inter-atomic interactions. Since $E$ and every $E[\varphi_a]$ (energies of atoms) are exact in all the above theories, the sum $\sum_{a<b} E^{\leftrightarrow}[\varphi_a\varphi_b]$ is also exact. By repeating this argument for parts of the system, we conclude that every term $E^{\leftrightarrow}[\varphi_a\varphi_b]$, and thus every inter-atomic XC term, is exact and universal to the lowest order in $\varphi_a\varphi_b$. As a demonstration of the asymptotic correspondence, it is found to $O(S^0)$ that $E_{xc}^{\infty\infty} = -(ab|ba)$ in $H_2$ in both HF/CI (Appendix A) and VB (Appendix B) theories. The NTB↔ HF/CI correspondence and the pairwise additivity of $E_{xc}^{\infty\infty}$ in HF/CI are justified in Appendices A and C, respectively.

*Asymptotic correspondence to $O(S^1)$*. To the next order in the inter-atomic overlap $S$, the above theories are no longer correspondent. This is evident from Table 1, where it is shown that the theories exhibit different $S$-dependent prefactors in $H_2$ inter-atomic energy terms (see Section S3 for derivations). This leads to the question about which theory is most compatible with NTB for the subsequent $E_{xc}^{\infty\infty}$ identification.

Table 1. Overlap-dependent prefactors in the inter-atomic energy terms in $H_2$.

| | $S$-dependent prefactor | Lowest order in $S$ |
|---|---|---|
| HF/CI | $(1+S)^{-2}$ | $-2S$ |
| VB | $(1+S^2)^{-1}$ | $-S^2$ |
| KS-DFT | $(1+S)^{-1}$ | $-S$ |
| 1-particle CI | $(1+S)^{-1}$ | $-S$ |

Analysis of the electron density expansion reveals that KS-DFT is the most suitable theory, asymptotically correspondent with NTB to $O(S)$. Considering a stretched $H_2$ molecule, its electron density in a minimal basis set (complete at $R \gg 0$) is $\rho(\mathbf{r}) = (1+S)^{-1}(\rho_a(\mathbf{r}) + \rho_b(\mathbf{r}) + 2\varphi_a^*(\mathbf{r})\varphi_b(\mathbf{r}))$, where $\rho_a$ and $\rho_b$ are defined in eq. (12). After linearizing it with respect to $S$ and $\varphi_a^*(\mathbf{r})\varphi_b(\mathbf{r})$ and neglecting $O(S^2)$ terms, we obtain

$$\rho = \rho_a + \rho_b + \delta\rho + O(S^2),$$
$$\delta\rho = -S(\rho_a + \rho_b) + 2\varphi_a^*\varphi_b. \qquad (41)$$

Since the $O(\delta\rho)$ errors in density lead to $O(\delta\rho^2)$ errors in total energy due to the variational principle, and since $\delta\rho \sim S \sim \varphi_a^*\varphi_b$, the neglect of the differential overlap-dependent terms in density shall lead to only $O(S^2)$ errors in energy. Thus, to $O(S)$, the KS-DFT total energy is a functional of the superposition of atomic densities $\rho = \rho_a + \rho_b$, similar to the NTB theory – it is said that two theories are $O(S)$-correspondent. As the HF/CI energy exhibits a different overlap dependence, NTB and HF/CI are only $O(S^0)$-correspondent.

The above arguments suggest the following procedure for obtaining mathematical forms of $E_{xc}^{\circ\!\!-\!\!\circ}$ and $\mu_{xc}^{\circ\!\!-\!\!\circ,ak}$ in NTB. KS-DFT equations are first expanded and truncated up to and including $O(S)$ terms. Then, the resulting KS-DFT inter-atomic terms are matched with $E_{xc}^{\circ\!\!-\!\!\circ}$ and $\mu_{xc}^{\circ\!\!-\!\!\circ,ak}$ of NTB. Finally, mathematical forms of the remaining integrals are obtained by comparison with HF/CI to $O(S^0)$.

## D. Atomion equation

The $O(S)$-correspondence of KS-DFT and NTB theories enables us to obtain insights into the terms contributing to $\mu_{xc}^{\circ\!\!-\!\!\circ,a}$ in eq. (30). This correspondence, implicitly assumed and utilized in Paper I, is now refined and generalized in Section S4 in the context of the atomion equalization principle (Section IVC), the asymptotic pairwise additivity of inter-atomic XC energies (Appendix C), and the systematic retention of $O(S)$ terms. Hereafter, the one-orbital convention is employed for simplicity, unless stated overwise – every atom is assumed to contribute only one atomion or atomic orbital to the system.

After $O(S)$ expansion of KS equations, the following *atomion equation* is obtained:

$$\varepsilon_a|\varphi_a\rangle = \left[-\frac{1}{2}\nabla^2 + v_{es}^a + \mu_{xc}^a + \sum_{b\neq a}\left(v_{es}^b + \mu_{xc,ba}^{R\gg 0} - P_b\mu_{xc,ba}^{R\gg 0}\right)\right]|\varphi_a\rangle, \qquad (42)$$

where $P_b = |\varphi_b\rangle\langle\varphi_b|$ is the projection operator; $\mu_{xc}^a$ was defined in eq. (29), and $\mu_{xc,ba}^{R\gg 0}$ is the asymptotic inter-atomic XC potential involving only two atoms: $a$ and $b$.

Comparison of eq. (42) and (28) reveals that the inter-atomic exchange-correlation potential is:

$$\mu_{xc}^{\circ\!\!-\!\!\circ,a}|\varphi_a\rangle = \sum_{b\neq a}\left(\mu_{xc,ba}^{R\gg 0} - |\varphi_b\rangle\langle\varphi_b|\mu_{xc,ba}^{R\gg 0}|\right)|\varphi_a\rangle\,. \qquad (43)$$

Eq. (42) bears strong resemblance to the Anderson equation,[49] introduced by P.W. Anderson in 1968,

$$\varepsilon_a|\varphi_a\rangle = \left[-\frac{1}{2}\nabla^2 + V_a + \sum_{b\neq a}(V_b - P_bV_b)\right]|\varphi_a\rangle. \qquad (44)$$

Eq. (42) generalizes the original Anderson equation to non-empirical potentials $V_b$ and has a firm theoretical basis in the adiabatic connection formalism, independent atom ansatz, and the asymptotic analysis. It has been successfully employed in Paper I.

Having established the form of $\mu_{xc}^{\circ\!\!-\!\!\circ,a}$, next, the inter-atomic IA-XC energy term $E_{xc}^{\circ\!\!-\!\!\circ}$ is discussed.

## E. Hückel electronic structure problem

Weeks, Anderson, and Davidson[210] demonstrated that rearrangement of eq. (44) leads to the Hückel (orthogonal tight-binding) electronic structure problem. By extending their approach to eq. (42)[183] (Section S6), the following standard eigenvalue problem is obtained:

$$\mathbf{D}\mathbf{c}_i = \varepsilon_i^{mo}\mathbf{c}_i, \qquad (45)$$

where $D_{aa}$ and $D_{ab}$ elements of the Hückel matrix $\mathbf{D}$ are defined as

$$D_{ba} = \langle\varphi_b|\mu_{xc,ba}^{R\gg 0}|\varphi_a\rangle \text{ for } b \neq a \qquad (46)$$

$$D_{aa} = \varepsilon_a = H_{aa} - \sum_{b\neq a} S_{ab}D_{ba}\,, \qquad (47)$$

$$H_{aa} = \langle\varphi_a|H_{KS,a}^{R\gg 0}|\varphi_a\rangle, \qquad (48)$$

and where $H_{KS,a}^{R\gg 0}$ is the asymptotic form of the KS Hamiltonian $H_{KS}$, defined by eq. (S23) in the Supporting Information. Throughout the article, $\varepsilon_i^{mo}$ are referred to as either Hückel energies or molecular orbital (MO) energies, and off-diagonal $D_{ab}$ – as resonance integrals.

Although the atomion equation and thus the Hückel problem have been derived in the $O(S)$ limit, it is remarkable that the Hückel energies $\varepsilon_i^{mo}$ and eigenvectors $\mathbf{c}_i$ are identical by construction to those of the KS generalized eigenvalue problem up to a normalization constant of $\mathbf{c}_i$:[210]

$$\mathbf{H}_{KS}^{R\gg0}\boldsymbol{c}_i = \varepsilon_i^{mo}\mathbf{S}\boldsymbol{c}_i, \tag{49}$$

where $(\mathbf{H}_{KS}^{R\gg0})_{ab} = \langle\varphi_a|H_{KS,a}^{R\gg0}|\varphi_b\rangle$ and $(\mathbf{S})_{ab} = \langle\varphi_a|\varphi_b\rangle$. This equivalence holds as long as $|\varphi_a\rangle$ states are the solutions to the atomion equation (eq. (42)). In other words, the derived equations are accurate to all orders in overlap ($O(S^\infty)$-accurate), at least for the asymptotic Hamiltonian $\mathbf{H}_{KS}^{R\gg0}$. In Section VI, it is argued that $\mathbf{H}_{KS}^{R\gg0}$ can be replaced with the full KS Hamiltonian $\mathbf{H}_{KS}$ for the self-consistent NTB theory, making the Hückel eigenvalues and eigenvectors formally exact. The connection between the $O(S)$ atomion equation and the $O(S^\infty)$ Hückel problem further justifies our approach of using asymptotics to obtain accurate mathematical forms of $E_{xc}^{\infty}$ and $\mu_{xc}^{\infty,a}$.

The $|\varphi_a\rangle$ states in eq. (45) appear orthogonal, the phenomenon which I refer to as *surprise orthogonality*. Their actual non-orthogonality is accounted for by overlap-dependent shifts $-S_{ab}D_{ba}$ of diagonal elements $D_{aa}$ in eq. (47). Such shifts have previously been used in theories of chemisorption,[206] the perturbational MO theory,[211] and semiempirical methods such as $\text{OM}_\text{x}$.[174, 212] Prior to these approaches, Harrison[104] and, earlier, Landshoff,[213] Wannier,[214] and Löwdin[98, 215] showed in the context of the empirical tight binding theory that the surprise orthogonality and the $D_{aa}$ shifts emerge if one defines new basis set functions $|\varphi_a'\rangle = |\varphi_a\rangle - \frac{1}{2}\sum_{b\neq a} S_{ba}\,|\varphi_b\rangle$ and ignores $O(S^2)$ terms. While the Harrison's, Löwdin's, and similar developments were limited to small overlaps, the Weeks-Anderson-Davidson's construction, originally proposed by Anderson and generalized in this work, is $O(S^\infty)$-accurate and is thus valid at any inter-atomic distances.

## F. Total energy expression in the nonempirical tight binding theory

Derivation of the total energy expression is nearly identical to the one reported in Paper I. It takes advantage of the atomic density additivity to $O(S)$ and the equivalence between Hückel and KS eigenvalues, described in Section VE. Its extension that accounts for charge transfer is presented in Section S7. The resulting expression is ($\rho_a$ is redefined as $|\varphi_a|^2$):

$$\begin{aligned} E = &\sum_a q_a \left\langle\varphi_a\middle|-\frac{1}{2}\nabla^2\middle|\varphi_a\right\rangle \\ &- \sum_{a,b\neq a} q_a S_{ab}\langle\varphi_b|\mu_{xc,ba}^{R\gg0}|\varphi_a\rangle \\ &+ \sum_{a,b\neq a} p_{ab}\langle\varphi_a|\mu_{xc,ab}^{R\gg0}|\varphi_b\rangle \\ &+E_{es}\left[\sum_a q_a\rho_a\right] + E_{xc}\left[\sum_a q_a\rho_a\right], \end{aligned} \tag{50}$$

where the bond order $p_{ab}$ is

$$p_{ab} = \sum_i f_i c_{ai}^* c_{bi}, \tag{51}$$

and $E_{es}$ was defined in eq. (23).

Eq. (50) can be written in a simplified form as

$$E = E_{kin} + E_{ortho} + E_{hyb} + E_{es} + E_{xc}, \tag{52}$$

where the kinetic energy $E_{kin}$, orthogonalization energy $E_{ortho}$, hybridization energy $E_{hyb}$, electrostatic energy $E_{es}$, and exchange-correlation energy $E_{xc}$ correspond to the respective terms in eq. (50) in the same order. The naming and physical significance of the $E_{hyb}$ and $E_{ortho}$ terms is in accord with the solid state physics literature.[206, 216] In Section S8, I show that the NTB total energy functional satisfies the force theorem.

Due to the asymptotic correspondence, eq. (50) and (20) must be identical in the large separation limit. Term-by-term comparison reveals that

$$E_{xc}^{\infty}[\rho] = E_{ortho} + E_{hyb} + E_{xc}^{\leftrightarrow}, \tag{53}$$

where $E_{xc}^{\leftrightarrow}$ is the inter-atomic XC functional in the KS (MO) ansatz, defined as

$$E_{xc}^{\leftrightarrow} = E_{xc}[\rho] - \sum_a E_{xc}^a[q_a\rho_a], \tag{54}$$

where $E_{xc}$ and $E_{xc}^a$ is the KS-XC energy of a molecule and atom $a$ *in a molecule*, respectively. I refer to $E_{xc}^{\leftrightarrow}$ as the *inter-atomic MO-XC functional,* which is distinct from the *inter-atomic IA-XC functional* $E_{xc}^{\infty}$. Notably, orthogonalization and hybridization energies are classified as inter-atomic XC effects in the NTB theory.

## VI. SELF-CONSISTENCY AND FORMAL EXACTNESS OF THE NON-EMPIRICAL TIGHT BINDING THEORY

As the NTB theory shall be self-consistent, the constrained variational minimization of the total energy expression (eq. (50)) shall yield the atomion equation (eq. (42)). The latter is derivable provided that the following equalities hold:

$$\frac{\delta E_{hyb}}{\delta \rho_a} = 0 \; \forall \, a, \tag{55}$$

$$\frac{\delta\left(S_{ab}\langle\varphi_b|\mu_{xc,ba}^{R\gg 0}|\varphi_a\rangle\right)}{\delta \rho_b} = 0 \; \forall \, b \neq a. \tag{56}$$

These equalities are justified by the fact that the expressions in numerators are not functionals of $\rho_a$ and $\rho_b$ densities, respectively – this subtle point is discussed in Section VIIC. Variational optimization of the total energy leads to the following self-consistent form of the atomion equation:

$$\varepsilon_a|\varphi_a\rangle = \left[-\frac{1}{2}\nabla^2 + v_{es}^a + \sum_{b\neq a} v_{es}^b + \mu_{xc}^a + \mu_{xc}^{\leftrightarrow,a} - \sum_{b\neq a} P_b \mu_{xc,ba}^{R\gg 0}\right]|\varphi_a\rangle, \tag{57}$$

where

$$\mu_{xc}^{\leftrightarrow,a} = \frac{1}{q_a}\frac{\delta E_{xc}^{\leftrightarrow}}{\delta \rho_a} \tag{58}$$

and

$$\mu_{xc}^{a} = \frac{1}{q_a}\frac{\delta E_{xc}^{a}}{\delta \rho_a}. \tag{59}$$

The original atomion equation (eq. (42)) is obtained if $\mu_{xc}^{\leftrightarrow,a} = \sum_{b\neq a}\mu_{xc,ab}^{R\gg 0}$ is assumed, which is only true in the $S \to 0$ limit. Although this pairwise additivity of $\mu_{xc}^{\leftrightarrow,a}$ was successfully used in Paper I, eq. (57) allows for more general cases when $E_{xc}^{\leftrightarrow}$ is not pairwise additive.

The NTB total energy functional, derived in a general form in Section III, is formally exact by construction. Notably, its $O(S)$ form (eq. (50)) is also formally exact even at large overlaps, provided that the exact $\mu_{xc,ab}^{R\gg 0}$ and $E_{xc}$ are known. To show this, I first recognize that

$$\sum_a E_{xc}^a + E_{xc}^{\leftrightarrow} = E_{xc}$$
$$\frac{1}{q_a}\frac{\partial E_{xc}}{\partial \rho_a} = \frac{\partial E_{xc}}{\partial \rho} = \mu_{xc}$$
$$\frac{\partial E_{xc}^b}{\partial \rho_a} = 0. \tag{60}$$

Therefore,

$$\mu_{xc}^a + \mu_{xc}^{\leftrightarrow,a} = \mu_{xc} \; \forall a \tag{61}$$

It follows that for the exact electron density, eq. (57) is equivalent to

$$\varepsilon_a|\varphi_a\rangle = \left[H_{KS} - \sum_{b\neq a} P_b \mu_{xc,ba}^{R\gg 0}\right]|\varphi_a\rangle. \tag{62}$$

If the Hückel theory derivation (Section S6) is repeated for such a form of the self-consistent atomion equation, the Hückel eigenvalues $\varepsilon_i^{mo}$ and eigenvectors $\mathbf{c}_i$ will be found to be identical to the solutions of $H_{KS}|\psi_i\rangle = \varepsilon_i^{mo}|\psi_i\rangle$. For the exact $E_{xc}$, they would coincide with the exact KS solutions. From the same analysis it would also follow that both $\varepsilon_i$ and $\mathbf{c}_i$ are invariant to the form of $\mu_{xc,ba}^{R\gg 0}$, provided that atomions are the solutions to eq. (57). The choice of $\mu_{xc,ba}^{R\gg 0}$ would, however, affect the shape of $|\varphi_a\rangle$ (cf. eq. (62)) and thus influence the electron density shape through eq. (11), (12), and (13):

$$\{\mu_{xc,ba}^{R\gg 0}\} \Rightarrow \{\varphi_a\} \Rightarrow \{q_a, \rho_a\} \Rightarrow \rho. \tag{63}$$

It then follows that there shall exist a set of $\{\mu_{xc,ba}^{R\gg 0}\}$ that produces the exact electron density of the system in the independent atom ansatz. The exact $\rho$ and $\varepsilon_i^{mo}$ shall yield the exact total energy through eq. (S46), which is equivalent to eq. (50). This concludes the proof that the NTB theory in the $O(S)$ form is formally exact.

The inter-atomic XC potential $\mu_{xc,ba}^{R\gg 0}$ is the essential component of the resonance integral $D_{ba} = \langle\varphi_b|\mu_{xc,ba}^{R\gg 0}|\varphi_a\rangle$ that determines $E_{hyb}$ and $E_{ortho}$ and is responsible for chemical bond formation. In the following section, I employ the asymptotic analysis to derive $D_{ba}$ and $\mu_{xc,ba}^{R\gg 0}$ mathematical forms. The forms will be shown to be identical to those that yielded accurate potential energy curves in Paper I.

## VII. MATHEMATICAL FORM OF THE RESONANCE INTEGRAL

The resonance integral $D_{ba}$ and the corresponding XC potential $\mu_{xc,ba}^{R\gg 0}$ are obtained using two complimentary methods: (1) through the analysis of the atomion equation in the $S \to 0$ limit, and (2) through the NTB↔HF/CI $O(S^0)$-asymptotic correspondence. One orbital per atom convention is used throughout.

### A. Resonance integral from the negligible overlap limit

In the $S \to 0$ limit, the overlap-free form of the atomion equation shall hold (eq. (S25)). Using the definition of $H_{KS,a}^{R\gg 0}$ in eq. (S23), writing eq. (S25) in the integral form, recognizing that $\langle\varphi_a|\mu_{xc,ba}^{R\gg 0}|\varphi_a\rangle = \langle\varphi_a|P\mu_{xc,ba}^{R\gg 0}|\varphi_a\rangle$, where $P = \sum_{i=1}^{N_{occ}}|\psi_i\rangle\langle\psi_i|$ is the projection operator, and carrying out constrained variational minimization, the following expression is obtained:

$$\varepsilon_a|\varphi_a\rangle = \left(H_{KS,a}^{R\gg 0} - \sum_{b\neq a} P\mu_{xc,ba}^{R\gg 0}\right)|\varphi_a\rangle. \tag{64}$$

After defining

$$H_{KS',a}^{R\gg 0} = H_{KS,a}^{R\gg 0} - \sum_{b\neq a}\mu_{xc,ba}^{R\gg 0}, \tag{65}$$

eq. (64) is rearranged as

$$\varepsilon_a|\varphi_a\rangle = \left[H_{KS',a}^{R\gg 0} + \sum_{b\neq a}\left(\mu_{xc,ba}^{R\gg 0} - P\mu_{xc,ba}^{R\gg 0}\right)\right]|\varphi_a\rangle, \tag{66}$$

Eq. (66) bears similarity to the Adams-Gilbert equation[47, 48] in its non-Hermitian[210] form. Since eq. (64) can also be written as $\varepsilon_a|\varphi_a\rangle = H_{KS',a}^{R\gg 0}|\varphi_a\rangle$, the following identity follows from eq. (66) that holds in the $S \to 0$ limit:

$$\left(\mu_{xc,ba}^{R\gg 0} - P\mu_{xc,ba}^{R\gg 0}\right)|\varphi_a\rangle = 0. \tag{67}$$

It is recognized that

$$P\mu_{xc,ba}^{R\gg 0}|\varphi_a\rangle = \sum_{d,e}^{M}|\varphi_d\rangle(\mathbf{S}^{-1})_{de}\langle\varphi_e|\mu_{xc,ba}^{R\gg 0}|\varphi_a\rangle. \tag{68}$$

Since $b \neq a$, all integrals $\langle\varphi_e|\mu_{xc,ba}^{R\gg 0}|\varphi_a\rangle$ are inter-atomic. At $R \to \infty$, the slowly varying $\varphi_a$ tails make leading contributions to $\langle\varphi_e|\mu_{xc,ba}^{R\gg 0}|\varphi_a\rangle$, and are denoted by $\varphi_a^\infty$. Since $\varphi_a^\infty$ is expected to cross over all neighboring atoms in the asymptotic limit, we have

$$\lim_{S\to 0}\langle\varphi_e|\mu_{xc,ba}^{R\gg 0}|\varphi_a\rangle = \lim_{\varphi_a^\infty\to 0}\langle\varphi_e|\mu_{xc,ba}^{R\gg 0}\rangle\varphi_a^\infty. \tag{69}$$

Then it follows from eq. (67), (68) and (69) that in the $S \to 0$ limit,

$$|\mu_{xc,ba}^{R\gg 0}\rangle - \sum_{d,e}^{M}|\varphi_d\rangle(\mathbf{S}^{-1})_{de}\langle\varphi_e|\mu_{xc,ba}^{R\gg 0}\rangle = 0. \tag{70}$$

Formally, any $\mu_{xc,ba}^{R\gg 0}$ operator of the form $|\mu_{xc,ba}^{R\gg 0}\rangle = \sum_c s_c|\varphi_c\rangle$ shall satisfy eq. (70), since $\sum_e(\mathbf{S}^{-1})_{de}\langle\varphi_e|\varphi_c\rangle = \delta_{dc}$. However, since by the $\mu_{xc,ba}^{R\gg 0}$ definition, only electrons belonging to atoms $a$ and $b$ contribute to it, it is only possible that $|\mu_{xc,ba}^{R\gg 0}\rangle = \alpha_a|\varphi_a\rangle + \alpha_b|\varphi_b\rangle$. Since a condition holds that $\langle\varphi_a|\mu_{xc,ba}^{R\gg 0}|\varphi_a\rangle \to 0$ at $S_{ab} \to 0$, $\alpha_a = 0$. It follows from the above analysis that the asymptotic *local* form of $\mu_{xc,ba}^{R\gg 0}$ has a particularly simple form:

$$\mu_{xc,ba}^{R\gg 0}(\mathbf{r}) = \alpha_b\varphi_b(\mathbf{r}) \tag{71}$$

where $\alpha_b$ is some (negative) coefficient. The obtained $\mu_{xc,ba}^{R\gg 0}$ is referred to as the *wave potential.* For many-orbital atoms, $\varphi_b$ should necessarily be an *s* orbital due to space isotropy. It should also be real and be taken with the positive sign to obtain real energies. The numerical value of $\alpha_b$ for $H_x$ systems is determined in Section VIIB. The wave potential gives rise to the resonance integrals of the form

$$D_{ba} = \alpha_b\langle\varphi_b|\varphi_b|\varphi_a\rangle, \tag{72}$$

which I refer to as *triple-orbital (or triple-O) resonance integrals.* In Section VIIC, I show that triple-O integrals corroborate inequalities in eq. (55) and (56) and are essential for the self-consistent theory. Their numerical accuracy has been assessed for model systems in Paper I.

Eq. (71) satisfies the following identity:

$$P|\mu_{xc,ba}^{R\gg0}\rangle = P_b|\mu_{xc,ba}^{R\gg0}\rangle. \quad (73)$$

This equation resembles the $PV_b = P_bV_b$ identity first postulated by Weeks et. al.[210] to connect Anderson and Adams-Gilbert equations, which was originally employed in Paper I to arrive at eq. (71). As it follows from the analysis presented herein, eq. (73) is only valid in the limit of small overlap.

### B. Resonance integral prefactor from the asymptotic correspondence principle

The $\alpha_b$ prefactor in the wave potential $\mu_{xc,ba}^{R\gg0}$ and in the resonance integral $D_{ba}$ can be obtained by leveraging the NTB↔HF/CI $O(S^0)$-asymptotic correspondence. In Section S4, it is shown that at large separations, only the degenerate valence atomions mix and contribute the leading terms to the energy expression. Therefore, it shall be sufficient to carry out the present analysis for $H_x$ structures – its findings are expected to be transferable to systems of many-electron atoms. Furthermore, in Appendix C it is shown that exchange and correlation in the large separation limit become pairwise-additive. Therefore, the $\alpha_b$ prefactor can be obtained by considering the simplest $H_2$ molecule.

The $O(S^0)$ asymptotic form of the $H_2$ inter-atomic XC energy in HF/CI is (Appendices A and B):

$$E_{xc}^{\circ\!\!-\!\!\circ}|_{HF/CI}^{R\gg0} = -\frac{1}{2}(ab|ba)|_{HF} - \frac{1}{2}(ab|ab)|_{CI} + O((ab)^4), \quad (74)$$

where the terms correspond to inter-atomic exchange, static correlation, and dynamic correlation, respectively.

In eq. (53), the two leading contributions to $E_{xc}^{\circ\!\!-\!\!\circ}$ that remain to $O(S^0)$ in NTB are the hybridization energy $E_{hyb}$ and the inter-atomic XC energy $E_{xc}^{\leftrightarrow}$. Consequently, the following *asymptotic correspondence identity* can be introduced:

$$E_{hyb} + E_{xc}^{\leftrightarrow} = E_x^{\circ\!\!-\!\!\circ}|_{HF}^{R\gg0} + E_c^{\circ\!\!-\!\!\circ}|_{CI}^{R\gg0} \quad (75)$$

To match the respective terms on the lefthand and righthand sides, the following arguments are considered: (1) $E_{hyb}$ and $E_{xc}^{\leftrightarrow}$ are connected through $\mu_{xc,ba}^{R\gg0}$ in the asymptotic limit; (2) $\langle a|\mu_x|a\rangle = (ab|ba) = E_x$; and (3) inter-atomic exchange and correlation are equal at $R \gg 0$. It is then concluded that $E_{hyb}$ can be matched with one of the terms in eq. (74), equal to $-0.5(ab|ba)$. Since $E_{hyb} = 2D_{ab}$ for $H_2$, the non-local precursor to the resonance integral takes a very simple form:

$$D_{ab} = \langle\varphi_a|\mu_{xc,ab}^{R\gg0}|\varphi_b\rangle = -\frac{1}{4}(ab|ab) = -\frac{1}{4}\int\frac{\varphi_a(\mathbf{r}')\varphi_b^*(\mathbf{r}')\varphi_a^*(\mathbf{r})\varphi_b(\mathbf{r})}{|\mathbf{r}-\mathbf{r}'|}d\mathbf{r}d\mathbf{r}'. \quad (76)$$

To obtain the local form of $D_{ab}$ (eq. (72)), we note that $\mu_{xc,ab}^{R\gg0}$ enters the atomion equation as $\mu_{xc,ba}^{R\gg0}|\varphi_a\rangle$ in the $S \to 0$ limit (see eq. (42)). Then, since $(ab|ab) = (ab|ba)$, we can write

$$\mu_{xc,ba}^{R\gg0}\varphi_a(\mathbf{r}) = -\frac{1}{4}\varphi_b(\mathbf{r})\int\frac{\varphi_b^*(\mathbf{r}')\varphi_a(\mathbf{r}')}{|\mathbf{r}-\mathbf{r}'|}d\mathbf{r}'. \quad (77)$$

The local form of eq. (77) is derived by adopting a spherical coordinate system, recognizing that $\varphi_a$ is slowly varying near $b$, and using the Laplace expansion for the $1/|\mathbf{r}-\mathbf{r}'|$ potential. After simplifications, the following result is obtained (see Section S9 for details):

$$\mu_{xc,ba}^{R\gg0}(\mathbf{r}) = -\frac{1}{2}\sqrt{\pi}\varphi_b(\mathbf{r})\int\frac{1}{r_>}R_b^*(r')(r')^2dr', \quad (78)$$

where $R_b^*$ is the radial part of $\varphi_b^*$.

In the $R \gg 0$ limit, the $\varphi_b^*(\mathbf{r}')\varphi_a(\mathbf{r}')$ product in eq. (77) will be so small in magnitude everywhere that only the integrand values with $\mathbf{r}$ being close to $\mathbf{r}'$ would make an appreciable contribution to the integral to make $1/|\mathbf{r}-\mathbf{r}'|$ large. Therefore, we can take $r_> \to r'$ in eq. (78) and obtain the integral $\int r'R_b^*(r')dr'$. For a 1s state $R_b^*(r')$, eq. (77) becomes

$$\mu_{xc,ba}^{R\gg0}(\mathbf{r}) = -\sqrt{\pi}\varphi_b(\mathbf{r}). \quad (79)$$

Comparison with eq. (71) shows that $\alpha_b = -\sqrt{\pi}$. Then, the resonance integrals are:

$$D_{ba} = -\sqrt{\pi}\langle\varphi_b|\varphi_b|\varphi_a\rangle. \quad (80)$$

The analysis above provides a systematic derivation of the $-\sqrt{\pi}$ prefactor in $D_{ba}$ integrals, introduced in a rather heuristic manner in Paper I.

## C. Self-consistency and interpretation of energy terms

*Orbital parity rule and void eigenpotential principle.* In Section VI, it was shown that the NTB theory is self-consistent provided that the following equalities hold:

$$\frac{\delta E_{hyb}}{\delta \rho_a} = 0 \;\forall\; a, \tag{81}$$

$$\frac{\delta\left(S_{ab}\langle\varphi_b|\mu_{xc,ba}^{R\gg0}|\varphi_a\rangle\right)}{\delta\rho_b} = 0 \;\forall\; b \neq a. \tag{82}$$

To justify eq. (81) and (82) using triple-O integrals, I propose a very simple *orbital parity rule*. In the $E_{ortho}$ terms such as $\langle\varphi_a|\varphi_b\rangle\langle\varphi_b|\varphi_b|\varphi_a\rangle$, there is an even number of $\varphi_a$ functions and an odd number of $\varphi_b$ functions. Since $\rho_a = |\varphi_a|^2$, both positive and negative values of $\varphi_a$ yield the same density $\rho_a$. Consequently, the term $\langle\varphi_a|\varphi_b\rangle\langle\varphi_b|\varphi_b|\varphi_a\rangle$ can adopt one (positive) value per $\rho_a$, but will take two (positive and negative) values per $\rho_b$, since its sign is determined by the sign of $\varphi_b$. Due to such duality, the term can no longer be regarded as a unique functional of $\rho_b$. We can then argue that the functional derivative with respect to $\rho_b$ cannot be justified, and eq. (82) holds. In contrast, the functional derivative with respect to $\rho_a$ is justified, leading to the corresponding term in the atomion equation (eq. (57)). Similarly, flipping the sign of either $\varphi_a$ or $\varphi_b$ will make $D_{ab} = -D_{ba}$ (for $H_2$) and thereby yield a complex $E_{hyb}$ value. Since both real and complex $E_{hyb}$ values correspond to the same $\rho_a$ (or $\rho_b$), $E_{hyb}$ is not a unique functional of either density, and its functional derivatives with respect to both $\rho_a$ and $\rho_b$ shall not exist. This leads to eq. (81) and corroborates the absence of $E_{hyb}$-derived terms in the potential operator of the atomion equation (eq. (57)). Generalization of this finding suggests that the lack of a potential term associated with any eigenvalue problem involving triple-O integrals – I refer to this observation as *the void eigenpotential principle.*

*Physical interpretation of NTB inter-atomic terms through self-consistency.* The void eigenpotential principle sheds light on the physical significance of NTB interatomic terms $E_{hyb}$ and $E_{xc}^{\leftrightarrow}$. In Section VIIB, we have used the asymptotic correspondence to equate $E_{xc}^{\leftrightarrow} + E_{hyb}$ from NTB and $E_x^{\circ\!\!-\!\!\circ}|_{HF} + E_c^{\circ\!\!-\!\!\circ}|_{CI}$ from HF/CI in the $S \to 0$ limit. Since $E_c^{\circ\!\!-\!\!\circ}|_{CI}$ arises from the eigenvalue problem while $E_x^{\circ\!\!-\!\!\circ}|_{HF}$ does not, the void eigenpotential principle shall apply to $E_c^{\circ\!\!-\!\!\circ}|_{CI}$ in a similar manner as to $E_{hyb}$. This motivates the following association: $E_{hyb} \equiv E_c^{\circ\!\!-\!\!\circ}|_{CI}$, which constitutes *the static correlation interpretation (SC-interpretation) of hybridization energy.* Then, $E_x^{\leftrightarrow} \equiv E_x^{\circ\!\!-\!\!\circ}|_{HF}$, and $E_c^{\leftrightarrow}$ is associated with dynamic correlation effects that are negligible in the $S \to 0$ limit.

Further justification of the SC-interpretation of $E_{hyb}$ follows from the fact that it yields the correct asymptotic form of the wave potential: $\mu_{xc,ba}^{R\gg0} = -\sqrt{\pi}\varphi_b$. To observe this, we consider the $H_x$ system as an example and note that in the $S \to 0$ limit, $\mu_{xc}^{\leftrightarrow,a} = \delta(E_{xc}^{\leftrightarrow})/\delta\rho_a = \delta(E_x^{\leftrightarrow})/\delta\rho_a$. In turn, $E_x^{\leftrightarrow} = E_x^{\circ\!\!-\!\!\circ}|_{HF} = \sum_{b>a} E_{x,ba}^{\circ\!\!-\!\!\circ}$, where $E_{x,ba}^{\circ\!\!-\!\!\circ} = -0.25[(ab|ba) + (ba|ab)]$. As shown in Section VIIB, its local, asymptotic form is $E_{x,ba}^{\circ\!\!-\!\!\circ} = -\sqrt{\pi}(aba) - \sqrt{\pi}(bab)$. Based on the orbital parity rule, $-\sqrt{\pi}(aba)$ is the only term that is a functional of $\rho_a$. Therefore, $\mu_{xc}^{\leftrightarrow,a} = \mu_x^{\leftrightarrow,a} = -\sqrt{\pi}\sum_{b\neq a}\varphi_b$. Since $\mu_{xc}^{\leftrightarrow,a} = \sum_{b\neq a}\mu_{xc,ba}^{R\gg0}$ in the same limit, $\mu_{xc,ba}^{R\gg0} = \mu_{x,ba}^{R\gg0} = -\sqrt{\pi}\varphi_b$.

The SC-interpretation of $E_{hyb}$ results in the $E_{xc}^{\leftrightarrow}$ form that mirrors properties of local XC functionals employed in KS-DFT. In both the SC-interpretation and local KS-DFT, exchange emerges as the dominant XC effect. Furthermore, $E_c^{\leftrightarrow}$ exclusively captures dynamic correlation effects, due to the single-determinant nature of KS-DFT. Lastly, $E_{xc}^{\leftrightarrow}$ is local in both theories. These similarities between NTB and KS-DFT are consistent with the fact that in NTB, $E_{xc}^{\leftrightarrow}$ corresponds to the inter-atomic part of KS-DFT XC functionals (see eq. (60)).

# VIII. DISCUSSION

*Comparison with Hückel theory.* The non-empirical tight binding theory (NTB) offers a rather simple, internally consistent, and formally exact framework for performing quantum mechanical calculations of chemically bonded systems. Fundamentally, it generalizes the Hückel electronic structure model by accounting for (1) the repulsive orbital overlap, (2) electrostatic and XC effects, and (3) environment-dependent atomic orbital relaxation. In NTB, mathematical forms of off-diagonal matrix elements $D_{ab}$ are defined in terms of inter-atomic XC potentials $\mu_{xc,ba}^{R\gg0}$ and are derived systematically by leveraging asymptotic correspondence among various electronic structure theories. As demonstrated numerically in Paper I for simple model systems, the obtained asymptotic functional forms can be transferable to typical chemical bonding distances. According to the discussion in Section VB, such transferability stems from the partial cancellation of repulsive electron-electron and attractive nuclear-electron inter-atomic interactions, which together constitute a rather small perturbation.

*Comparison with other TB methods.* Although tight-binding (TB) methods are often regarded as approximations to KS-DFT[217] or HF[118] theories, it has also been recognized[218] that the TB theories shall be derivable from a different starting point, owing to the correct bond dissociation behavior exhibited by at least some of them. The author's opinion is that the independent atom ansatz, introduced in this work, provides such the starting point.

It is notable that NTB shares many features of successful TB models, such as (1) atomic density additivity,[158, 162, 217, 219] (2) pairwise inter-atomic exchange-correlation,[137, 158, 217, 219] (3) pairwise repulsive $S_{ab}D_{ba}$ terms,[104, 174, 206] (4) orthogonal eigenvalue problem,[118, 125, 174] (5) charge transfer decoupled from hybridization[217] that brings about matrix diagonal element shifts[137, 220] (see Section IVC), and (6) diatomic off-diagonal matrix elements.[158, 171, 217, 219] On the one hand, such similarity can be interpreted as a justification for why many simple TB models work so well. On the other hand, it may hint at the possible origin of their limitations. The superior performance of NTB over other TB models for model systems, as described in Paper I, can be attributed to the fact that not a single TB theory shares all six features with NTB. Among TB models, $OM_x$ methods[174, 212] stand apart as they share properties #3, #4, and #6 with NTB that affect leading orthogonalization and hybridization energy terms. This may explain superior accuracy of $OM_x$ over other TB methods.[181, 221]

*NTB hierarchical structure and expected computational gains.* A unique attribute of NTB is its hierarchical and multiscale structure. Minimization of various energy terms occurs at up to three different levels. There are atomions, obtained for the specific atomic environment by solving the corresponding eigenvalue problem. Atomions are mixed to form molecular orbitals by solving the Hückel problem, while atomion occupancies are optimized through the electronegativity equalization. Finally, occupied and virtual MOs may further mix in the CI treatment of the correlation energy, as will be discussed in Section IX. This hierarchical structure should considerably decrease the computational cost of QM calculations. As discussed in Paper I, the construction, orthogonalization, and diagonalization of the $(n_\zeta n_{val} M) \times (n_\zeta n_{val} M)$ matrix in KS-DFT, where $M$, $n_{val}$, and $n_\zeta$ are the number of identical atoms, valence atomic orbitals, and basis set functions per atomic orbital, respectively, is reduced to diagonalization of $M$ small $(n_{val} n_\zeta) \times (n_{val} n_\zeta)$ matrices associated with the atomion equation and of one large $(n_{val} M) \times (n_{val} M)$ matrix in the Hückel eigenvalue problem. NTB thus reduces the complete-basis-set electronic structure problem to the one involving the minimal basis set, optimized on the fly as a function of the environment. Notably, the cost of diagonalizing the $(n_{val} M) \times (n_{val} M)$ Hückel matrix shall also be lower in comparison with KS-DFT, as the apparent "surprise-orthogonality" of atomions (see Section VE) would eliminate a computationally intensive basis set orthogonalization step, while the analytical resonance integrals would replace numerical integration. Finally, interpretation of electronic structure calculations is considerably simplified in NTB, as unique atomic charges and energy contributions are obtained at no extra computational cost.

*Atom-molecule duality and topology.* Owing to its hierarchical structure, molecular systems described by NTB would exhibit both atomic (atomions) and molecular (MOs) features. While MOs are experimentally observed in real molecules,[222] atomions corroborate the notion of transferable "atoms in a molecule" that is so useful in analyses of chemical reaction mechanisms. Thus, NTB can be regarded as a candidate for the universal theory of localized orbitals, hypothesized in Section I.

The appearance of the Hückel electronic structure problem, deeply connected to topology,[223] in NTB would corroborate the crucial role that topology plays in describing chemical reactivity. Topological control of reactivity and energetics has been an integral part of a variety of methods and concepts, such as group additivity,[15, 16] cluster expansion,[224] linear scaling relationships,[225] coordination and generalized coordination numbers,[22] contributions of non-local topological features,[20] and the chemical graph theory in general.[226]

*Self-interaction and static correlation errors.* As NTB is formally exact, the question arises about the mechanisms for eliminating static correlation, one-electron and many-electron self-interaction errors that are responsible for many deficiencies of traditional KS-DFT methods.

The static correlation error (SCE) has been attributed to spurious electrostatic interaction between fractional spin-up and spin-down electrons on atoms.[227] It is responsible for the incorrect dissociation of chemical bonds[228] in the restricted HF or KS-DFT formalism. NTB is trivially free of SCE due to the natural separation of intra-atomic and inter-atomic electrostatic terms and the exact treatment of the onsite static correlation energy (see Appendix A).

One-electron self-interaction errors (1-SIE) arise when fractionally occupied KS states exhibit non-$f^2$ scaling of exchange energy, where $f$ is the MO occupancy, so that the electron self-interaction error is not exactly cancelled by self-exchange.[229] 1-SIE is a common problem of KS-DFT methods, as the exchange energy that is based on the homogeneous electron gas model scales as $f^{4/3}$.[207] In NTB, the intra-atomic 1-SIE is absent, as the onsite self-exchange is treated exactly (see Appendix A). The inter-atomic 1-SIE is absent by construction, as the inter-atomic electrostatic terms of the form $q_{be}q_{ak}\mathcal{F}_{beak}$, where $q_{be}$ and $q_{ak}$ are atomion charges and $\mathcal{F}_{beak}$ is defined in eq. (37), correspond to the physically present electron-electron repulsion and do not contain any spurious contributions. As 1-SIE is responsible for underestimation of reaction barriers,[230] NTB shall be able to predict accurate barriers, which has been demonstrated for the $2H_2 + D_2 \rightarrow 2HD + H_2$ reaction in Paper I.

Finally, I provide arguments to indicate that NTB is free of many-electron SIE (many-SIE). Many-SIE appears when total energy is non-linear with respect to addition and removal of fractional electron numbers.[231] To illustrate one possible mechanism for its removal in NTB, I consider again an $H_2$ molecule. Upon addition of $2\delta$ electrons, the kinetic, orthogonalization, hybridization, nuclear-electron electrostatic, and correlation energies shall change linearly. The inter-atomic electrostatic energy will, however, contain the repulsive term proportional to $\delta^2$, as each atom will acquire $\delta$ fractional electrons. Upon application of the atomion equalization principle and performing energy minimization with respect to atomion occupancies, however, the quadratic term will disappear, as it is more energetically favorable to localize $2\delta$ electrons on one atom. The symmetry of the molecule can then be restored if the final state is recognized as a superposition of two degenerate, non-coupled states each containing extra $2\delta$ electrons on atom 1 and atom 2, respectively. As a result of electron localization, the inter-atomic electrostatic energy will change linearly. Overall, the total energy shall change linearly with the $2\delta$ electron addition, rendering NTB many-SIE-free.

# IX. RE-INTERPRETATIONS IMPOSED BY PHYSICAL CONSTRAINTS

## A. Quantum fluctuations

Since the NTB theory development deals with analyzing exceedingly small inter-atomic terms in the large separation limit, it is essential to account for all terms of comparable magnitude. In Section VIIB, the argument was made that in the integral $D_{ab} = -0.25 \int \varphi_a(\mathbf{r}')\varphi_b^*(\mathbf{r}')\varphi_a^*(\mathbf{r})\varphi_b(\mathbf{r})/|\mathbf{r}-\mathbf{r}'|\, d\mathbf{r}d\mathbf{r}'$, when $R \gg 0$, only the $\mathbf{r}$ values approaching $\mathbf{r}'$ contribute significantly. However, at $|\mathbf{r}-\mathbf{r}'|$ distances approaching $\sim\hbar/mc$ ($\approx 1/137$ Bohr), quantum field effects become notable.[232] These effects are associated with the zitterbewegung – rapid quantum oscillations – and are not captured by the Schrödinger equation. Instead, they arise in a form of a Darwin term[233] from the relativistic Dirac equation.[234] The Darwin term acts to reduce the effect of an electrostatic potential, and for an H atom with the nucleus at $\mathbf{R}$ position, it takes the form $\Delta E_{Darwin} = \kappa|\varphi(\mathbf{R})|^2$, where $\kappa$ is a very small parameter. By analogy, this form can be adopted to alter inter-atomic IA-XC electron-electron integrals in $E_{xc}^{\circ\!-\!\circ}$ at $R \gg 0$ and be written as $\langle\varphi_b|\delta|\varphi_a\rangle$, where $\delta > 0$ is a small, unspecified constant. The $\langle\varphi_b|\delta|\varphi_a\rangle$ terms serve to reduce the magnitude of resonance integrals slightly, which is consistent with slight weakening of chemical bonds by relativistic effects, as has been observed, for example, in $H_2$,[235] HF, HCl, HBr, and HI.[236]

In NTB, the quantum effects are accounted for in the $R \gg 0$ limit through the following modifications:

$$\mu_{x,ba}^{R\gg0} \rightarrow \mu_{x,ba}^{R\gg0} + \delta, \tag{83}$$

$$D_{ba} = \langle\varphi_b|\mu_{x,ba}^{R\gg0}|\varphi_a\rangle \rightarrow \langle\varphi_b|\mu_{x,ba}^{R\gg0} + \delta|\varphi_a\rangle. \tag{84}$$

The effect of $\delta$ amounts to making $\mu_{x,ba}^{R\gg0}$ less negative. The modified form of $D_{ba}$ should be understood as involving $(\mu_{x,ba}^{R\gg0} + \delta)\Theta(-\mu_{x,ba}^{R\gg0} - \delta)$, where $\Theta$ is the Heaviside function, introduced to ensure that $D_{ba} \leq 0$ $\forall\, R$ and not shown in eq. (84) to simplify the notation. The term $\langle\varphi_b|\delta|\varphi_a\rangle$ can be regarded as the lowest energy scale at which the non-relativistic NTB theory breaks down. Effects associated with $\delta$ are very small at chemical bonding distances and can be neglected in actual calculations. However, they are essential for the asymptotic analysis at $R \gg 0$.

There is another physical effect that can be captured through the inclusion of $\delta$. If the resonance integral $D_{ba}$ is interpreted as the likelihood of interstate transitions, small integral values necessitate observations of the system over long time periods $\Delta t$. If the system is observed over $\Delta t' \ll \Delta t$, there is a high probability that the effects associated with $D_{ba}$ will not be observed. We can argue that the observation time interval $\Delta t'$ provides an extra degree of freedom that enables "tuning" or "turning off" $D_{ba}$ through $\delta$.

## B. Re-interpretation of hybridization energy and transition to the distinguishable particle limit

In this subsection, I demonstrate how accounting for quantum fluctuations necessitates a revision of the hybridization energy interpretation, paving the way toward more accurate $D_{ab}$ and $E_{xc}^{\leftrightarrow}$ approximations. Considering the $H_2$ molecule, we note that the $E_{hyb}$ derivative cannot be generally taken, since both complex and real values of $E_{hyb}$ correspond to the same density $\rho_a$ (Section VIIC). However, at a certain $R = R_\infty$, $\langle \varphi_b | \mu_{x,ba}^{R\gg 0} | \varphi_a \rangle = -\langle \varphi_b | \delta | \varphi_a \rangle$, $D_{ab} = D_{ba} = 0$, and $E_{hyb} = 0$. In this scenario, $E_{hyb}$ becomes invariant to the sign of $\varphi_a$, allowing us to take a variational derivative of it, since now it is a unique functional of $\rho_a$. Applying the orbital parity rule to $D_{ab}$ and $D_{ba}$, we find that, since the first part of $D_{ab} = \langle \varphi_a | -\sqrt{\pi}\varphi_a + \delta | \varphi_b \rangle$ contains two $\varphi_a$ atomions but one $\varphi_b$, it is a functional of $\rho_a$ but not $\rho_b$ (and vice versa for $D_{ba}$). Similarly, $\langle \varphi_b | \delta | \varphi_a \rangle$ is not a functional of any density. Using eq. (58), it then follows that $E_{hyb}$ contributes a $-\sqrt{\pi}\varphi_b$ term into $\mu_{xc}^{\leftrightarrow,a}$ and thus to $\mu_{x,ba}^{R\gg 0}$. However, an identical term has already been contributed by $E_x^{\leftrightarrow}$, leading to an extra factor of 2 in eq. (80). This, however, violates both the asymptotic correspondence and self-consistency within the SC-interpretation of $E_{hyb}$.

To resolve this, I propose *the exchange-static-correlation interpretation (XSC-interpretation)* of the hybridization energy. In this interpretation, the asymptotic identity (eq. (75)) $E_{xc}^{\leftrightarrow} + E_{hyb} = E_x^{\circ\!\!-\!\!\circ}|_{HF} + E_c^{\circ\!\!-\!\!\circ}|_{CI}$ is employed at distances sufficiently small to describe $E_c^{\circ\!\!-\!\!\circ}|_{CI}$ as a sum of static and dynamic correlation:

$$E_c^{\circ\!\!-\!\!\circ}|_{CI} = E_{c-s}^{\circ\!\!-\!\!\circ}|_{CI} + E_{c-d}^{\circ\!\!-\!\!\circ}|_{CI}, \qquad (85)$$

where $E_{c-s}^{\circ\!\!-\!\!\circ}|_{CI} = -0.5(ab|ba)$. Then, $E_{hyb}$ is matched with $E_x^{\circ\!\!-\!\!\circ}|_{HF} + E_{c-s}^{\circ\!\!-\!\!\circ}|_{CI}$, $E_x^{\leftrightarrow}$ is set to zero, and $E_c^{\leftrightarrow}$ is matched with $E_{c-d}^{\circ\!\!-\!\!\circ}|_{CI}$, in accord with its dynamic correlation interpretation.

In the XSC-interpretation, the resonance integrals are expressed as follows:

$$\begin{aligned} D_{ab} &= D_{ab}^{x} + D_{ab}^{sc}, \\ D_{ab}^{x} &= \langle \varphi_a | \mu_{x,ab}^{R\gg 0} + \delta | \varphi_b \rangle, \\ D_{ab}^{sc} &= \langle \varphi_a | \eta_{sc,ab}^{R\gg 0} | \varphi_b \rangle, \end{aligned} \qquad (86)$$

where $D_{ab}^{x}$ and $D_{ab}^{sc}$ are the exchange and static correlation contributions to $D_{ab}$ that yield the corresponding $E_{x-hyb}$ and $E_{sc-hyb}$ terms in $E_{hyb}$. In the $D_{ab}^{sc}$ expression, $\mu_{c,ab}^{R\gg 0}$ was replaced with $\eta_{sc,ab}^{R\gg 0}$ to account for the fact that any XC potential $\mu_{xc}$ must be associated with a single-determinant theory, and thus the static correlation energy shall not yield any corresponding $\mu_c$ term. The $\delta$ term was not included in $D_{ab}^{sc}$ to ensure self-consistency and the correct $\mu_{x,ba}^{R\gg 0} = -\sqrt{\pi}\varphi_b$ form. Heuristically, the neglect of $\delta$ can be understood by recognizing that static correlation involves electron hopping between degenerate configurations H↑/H↓ and H↓/H↑. Using the uncertainty relation $\Delta E \Delta t \sim \hbar$, we find that for $\Delta E \to 0$, $\Delta t \to \infty$ and thus $\Delta x \to \infty$ due to electron velocity limited by the speed of light, where $\Delta t$ and $\Delta x$ are the duration of observation of the system and the uncertainty in electron coordinates, respectively. As $\Delta x \gg \hbar/mc$, quantum fluctuations are not expected to play a role in $E_{c-s}^{\circ\!\!-\!\!\circ}|_{CI}$ and $D_{ab}^{sc}$.

A notable consequence of including quantum fluctuations in NTB is the mechanism governing the transition from electrons being delocalized and indistinguishable to becoming localized and distinguishable as a molecule undergoes dissociation to individual atoms. The exchange interaction, arising from the indistinguishability of particles,[237] must evidently vanish in the "distinguishable particle" limit. This occurs precisely at $R = R_\infty$, where $\langle \varphi_b | \mu_{x,ba}^{R\gg 0} | \varphi_a \rangle = -\langle \varphi_b | \delta | \varphi_a \rangle$, $D_{ab}^{x} = 0$, and thus $E_{x-hyb} = E_x^{\circ\!\!-\!\!\circ}|_{HF} = 0$. Since NTB orbital mixing ceases to exist in this limit, electrons become localized on atoms, and atomic spin occupancies transition from $H(\frac{1}{2};\frac{1}{2})$ to $H(1;0)$ and $H(0;1)$. This transition is evidently associated with large $\Delta E$ at small chemical bonding $R$ values, indicating that $\Delta t$ and thus $\Delta x$ are small (*vide supra*) and influenced by quantum fluctuations, supporting the inclusion of $\delta$ in $D^{x}_{ab}$. In contrast to other electronic structure theories, NTB with quantum fluctuations allows for the "indistinguishable $\rightarrow$ distinguishable" transition to occur at finite $R$ and observation times $\Delta t$.

**C. Revised self-consistent atomion equation**

The XSC-interpretation of $E_{hyb}$ requires a slight revision of the atomion equation (eq. (57)). We write it as

$$\varepsilon_a|\varphi_a\rangle = \left[-\frac{1}{2}\nabla^2 + v^a_{es} + \sum_{b\neq a} v^b_{es} + \mu^a_{xc} + \mu^{\leftrightarrow,a}_{xc} - \sum_{b\neq a} P_b\left(\eta^{R\gg 0}_{sc,ba} + \mu^{R\gg 0}_{x,ba} + \delta\right)\right]|\varphi_a\rangle, \quad (87)$$

where the $\mu^{\leftrightarrow,a}_{xc}$ potential is redefined as

$$\mu^{\leftrightarrow,a}_{xc} = \frac{\delta\left(E_{hyb} + E^{\leftrightarrow}_{c}\right)}{q_a\delta\rho_a}. \quad (88)$$

As described above, at a certain $R = R_\infty$, the derivative of the exchange part of $E_{hyb}$ can be taken, yielding the correct asymptotic potential $\mu^{R\gg 0}_{x,ba} = -\sqrt{\pi}\varphi_b$.

Concluding this subsection, I would like to point out one curious corollary to the above analysis. It was found that $\mu^{R\gg 0}_{x,ba} = -\sqrt{\pi}\varphi_b$ is a functional of the inter-atomic IA-XC functional only at the very specific condition – when the equality $\langle\varphi_b|\mu^{R\gg 0}_{x,ba}|\varphi_a\rangle = -\langle\varphi_b|\delta|\varphi_a\rangle$ is satisfied at $R_\infty$. At any smaller distance, the form of $\mu^{R\gg 0}_{x,ba}$ can no longer be obtained, since $E_{x-hyb}$ becomes non-differentiable. It follows that the identity $\mu^{\leftrightarrow,a}_{x} = \delta E_{x-hyb}/\delta\rho_a$ holds only at $R_\infty$, and the corresponding peculiar potential can be referred to as the *detached potential*, rendering NTB self-consistent only at $R = R_\infty$. The utility of the detached potential concept has been already demonstrated in Paper I, where the pairwise-additive asymptotic form $\mu^{R\gg 0}_{x,ba} = -\sqrt{\pi}\varphi_b$ that led to numerically accurate predictions, was not equal to the XC potential of the PBE functional used, but instead corresponded to the asymptotic form of the exact inter-atomic exchange, described in Section VII.

To explain the successful use of the detached potential, we must imagine a process in which we first bring atoms to nearly infinity where $E_{x-hyb} = 0$ identically, then take the functional derivative there, bring atoms back to the chemical bonding distances while retaining the same functional form of $\mu^{R\gg 0}_{x,ba}$, and only then evaluate the terms in the energy expression. In other words, this suggests employing the *translatio ex infinitum* technique not just for inter-atomic XC energy terms, but also for inter-atomic XC potentials. The remarkable decoupling of $E_{x-hyb}$ and $\mu^{\leftrightarrow,a}_{xc}$ at $R < R_\infty$ will be further illustrated numerically using the non-local form of the resonance integral (*vide infra*), which will be the topic of the forthcoming publication.

**D. Impact of further fundamental constraints**

In this section, I discuss two fundamental constraints that make a case for (1) full CI treatment of the dynamic correlation within NTB and (2) non-local resonance integrals $D_{ab}$ at chemical bonding distances.

*Constraint no. 1.* NTB must be consistent with VB theory. At $R = R_\infty$, inter-atomic exchange disappears, and the remaining static correlation yields $E^{\infty}_{xc} = -0.5(ab|ba)$. In VB theory, however, $E^{\infty}_{xc} = -(ab|ba)$. The NTB$\leftrightarrow$VB correspondence can be restored if $E^{\leftrightarrow}_{c}$ is constructed to yield another $-0.5(ab|ba)$ term at $R = R_\infty$. This extra contribution can be obtained if the dynamic correlation $E^{\leftrightarrow}_{c}$ follows from the CI eigenvalue problem with $-0.5(ab|ba)$ taken as an off-diagonal matrix element. Then, at large separations, $E^{\leftrightarrow}_{c}$ is expected to reduce to the static correlation, identical to $E_{sc-hyb}$, such that $E^{\leftrightarrow}_{c} + E_{sc-hyb} = -(ab|ba)$ at $R = R_\infty$. At chemical bonding $R$, the gap $\Delta E$ between ground and excited MOs is large, making $\Delta t$ and thus $\Delta x$ small and sensitive to quantum fluctuations. It follows that $E^{\leftrightarrow}_{c}$ must contain $\delta$ and thus contribute to $\mu^{\leftrightarrow,a}_{xc}$ in eq. (88) at $R = R_\infty$.

*Constraint no. 2.* The total NTB energy must be of $O((\varphi_a\varphi_b)^2)$. This follows from the DFT variational principle, according to which $\delta E \sim (\delta\rho)^2$, and since $\delta\rho \sim \varphi_a\varphi_b$ in KS-DFT at small overlaps (cf. Section VC). Notably, the local form of the resonance integral $D_{ba} = -\sqrt{\pi}\langle\varphi_b|\varphi_b|\varphi_a\rangle$ does not satisfy this constraint, since $D_{ba}$ and thus $\delta E \sim \varphi_a\varphi_b$, which may be the cause of slight overbinding at $R > R_0$, where $R_0$ corresponds to an energy minimum, as observed in Paper I. However, the non-local precursor of the resonance integral $D^{ex}_{ba} = -0.25(ba|ab)$ is of $O((\varphi_a\varphi_b)^2)$ and does satisfy this constraint. This observation suggests a modification of the *translatio ex infinitum* technique, where the asymptotic forms of both energy and potential are retained at chemical bonding distances, while local integrals are replaced with their non-local counterparts. In particular, the resonance integral $D_{ab} = D_{ab}^{x} + D_{ab}^{sc}$ would take the following form:

$$D_{ab} = -\frac{1}{2}(ab|ab)$$
$$= -\frac{1}{2}\int \frac{\varphi_a(\mathbf{r}')\varphi_b^*(\mathbf{r}')\varphi_a^*(\mathbf{r})\varphi_b(\mathbf{r})}{|\mathbf{r}-\mathbf{r}'|} d\mathbf{r}d\mathbf{r}'. \quad (89)$$

The non-local form of $D_{ab}$ has an important advantage over the local form – it ensures that the Hückel matrix **D** is always Hermitian.

## X. CONCLUSIONS

This article introduces a self-consistent, non-empirical framework for tight binding theory. The formalism is based on a new DFT ansatz, termed the independent atom ansatz, that expresses electron density in terms of non-interacting perturbed atomic orbitals. The ansatz takes advantage of the partial cancellation of inter-atomic repulsive electron-electron and attractive electron-nuclear interactions to enable rapid convergence of total energy with respect to interatomic overlap. Analytical expressions for inter-atomic terms are obtained using the asymptotic correspondence between formally exact electronic structure theories at large interatomic separations. The derived expressions were found to be numerically accurate for simple model systems in the previous work.[183] The non-empirical tight binding theory incorporates all physical effects that contribute to chemical bonding and offers mechanisms for the elimination of self-interaction and static correlation errors.

## ACKNOWLEDGEMENTS

The author thanks Alanna "Lanie" Leung for numerous valuable discussions and numerical tests (especially of equations in Chapter IX) that will constitute a forthcoming publication. I would also like to thank So Hirata and Luke Nambi Mohanam for the feedback on the manuscript and Baron Peters for proposing the Weeks-Chandler-Andersen analogy. This work was supported by the National Science Foundation under award number CHE-2154781.

## APPENDIX A

In this Appendix, the inter-atomic form of the HF/CI exchange-correlation energy in the $H_2$ molecule is obtained to $O(S^0)$. In the HF theory, the exchange energy has the following general form:[228]

$$E_x|_{HF} = \frac{1}{2}\sum_{\mu}\sum_{\nu}\left[P_{\nu\mu}^{\alpha}F_{\mu\nu}^{\alpha} + P_{\nu\mu}^{\beta}F_{\mu\nu}^{\beta}\right], \quad (90)$$

where $P_{\nu\mu}^{\alpha}$ and $P_{\nu\mu}^{\beta}$ are the $\alpha$-spin and $\beta$-spin density matrices, and $F_{\mu\nu}^{i}$ are the standard Fock matrix elements.

At large inter-atomic separations, the density matrix becomes diagonal. For atomic orbitals that become stationary states in the large separation limit (see Section VD), the following identity holds:[238]

$$\frac{\partial P_{\mu\nu}}{\partial t} = \frac{i}{\hbar}(\varepsilon_\nu - \varepsilon_\mu)P_{\mu\nu}$$
$$= \frac{i}{\hbar}\sum_{\sigma}(P_{\mu\sigma}H_{\sigma\nu} - H_{\mu\sigma}P_{\sigma\nu}), \quad (91)$$

where $H_{\mu\sigma}$ are the elements of the diagonal matrix of the Hamiltonian operator (such as in eq. (42)) in the representation of atomic states, $\varepsilon_\nu$ and $\varepsilon_\mu$ are the atomic orbital energies, and $t$ is time. Atomic orbitals are considered non-degenerate by $\delta\varepsilon \sim \hbar/\Delta t$, where $\Delta t$ is taken as large and finite. Evidently, $\frac{\partial P_{\mu\nu}}{\partial t} = 0$ in the ground state for the given $R$, **P** commutes with **H**, and **P** is thus diagonal.

For the minimal-basis $H_2$ with $P_{aa}^{\alpha} = P_{aa}^{\beta} = P_{bb}^{\alpha} = P_{bb}^{\beta} = 0.5$, eq. (91) becomes

$$E_x|_{HF}^{R\gg 0} = -\frac{1}{4}(aa|aa) - \frac{1}{4}(bb|bb)$$
$$-\frac{1}{2}(ab|ba), \quad (92)$$

where the first two terms are the intra-atomic self-exchange terms that remove the electron self-interaction error in the atomic limit.

The correlation energy in the large separation limit equals[228]

$$E_c|_{CI}^{R\gg 0} = -(\psi_1\psi_2|\psi_1\psi_2) \quad (93)$$

where $\psi_1 = \frac{1}{\sqrt{2}}(\varphi_a + \varphi_b)$ and $\psi_2 = \frac{1}{\sqrt{2}}(\varphi_a - \varphi_b)$ are the occupied and virtual molecular orbitals. For the diagonal density matrix, it turns out that $E_c = E_x$ in the large separation limit. Like $E_x^{R\gg 0}$, $E_c^{R\gg 0}$ contains terms $-\frac{1}{4}(aa|aa)$ and $-\frac{1}{4}(bb|bb)$ that eliminate intra-atomic static correlation error.[239] The inter-atomic XC part is then

$$E_{xc}^{\infty\!\!-\!\!\infty}|_{HF/CI}^{R\gg 0} = -(ab|ba). \tag{94}$$

Evidently, the kinetic and electrostatic energy terms in HF for the diagonal density matrix at $R \gg 0$ would coincide with the corresponding NTB terms. Thus, the asymptotic correspondence between NTB and HF/CI holds to $O(S^0)$ and $O((\varphi_a\varphi_b)^2)$, and $E_{xc}^{\infty\!\!-\!\!\infty} = E_{xc}^{\infty\!\!-\!\!\infty}|_{HF/CI}^{R\gg 0}$. In Appendix C, it is shown that $E_{xc}^{\infty\!\!-\!\!\infty}|_{HF/CI}^{R\gg 0}$ becomes pairwise-additive at $R \gg 0$. Therefore, the findings for the $H_2$ molecule shall be generalizable to systems with $> 2$ atoms.

## APPENDIX B

In this Appendix, the asymptotic, $O(S^0)$ form of the valence bond (Heitler-London) energy expression is derived for $H_2$. In the minimal-basis valence bond theory, energies of ground and excited states relative to non-interacting atoms are written as[240]

$$\Delta E_{\pm} = \frac{J \pm K}{1 \pm S^2}. \tag{95}$$

Here, " + " corresponds to the lowest-energy ground state; $J$ is the Coulomb integral, identical to the inter-atomic electrostatic energy $E_{es}^{\infty\!\!-\!\!\infty}$, defined in eq. (22); $S$ is the standard overlap integral; and $K$ is the exchange integral that can be expressed as

$$K = (\varphi_a\varphi_b|\varphi_b\varphi_a) + \frac{S^2}{R}$$
$$+S[(\varphi_a v_a \varphi_b) + (\varphi_b v_b \varphi_a)]. \tag{96}$$

To obtain the limiting form of $\Delta E_{\pm}$, valid to $O(S^0)$, we set $S = 0$ and find that

$$\Delta E_{\pm} = J \pm (\varphi_a\varphi_b|\varphi_b\varphi_a). \tag{97}$$

Curiously, now the " − " sign corresponds to the lowest energy state. Evidently, the last term corresponds to the inter-atomic correlation, and we conclude that $E_{xc}^{\infty\!\!-\!\!\infty} = E_c^{\infty\!\!-\!\!\infty}|_{VB} = -(\varphi_a\varphi_b|\varphi_b\varphi_a)$. Notably, the same expression is obtained if we construct a (2x2) CI matrix using atom-localized electrons as a reference state.

## APPENDIX C

In this Appendix, the pairwise additivity of the XC energy is shown at large inter-atomic separations. By invoking the asymptotic correspondence between HF/CI and KS-DFT to $O(S^0)$ (see Section VC), it will be sufficient to show the pairwise additivity of inter-atomic exchange and correlation in HF/CI.

*Pairwise-additive inter-atomic exchange.* As described in Appendix A, the asymptotic form of the exchange energy expression is

$$E_x|_{HF}^{R\gg 0} = -\frac{1}{2}\sum_{\mu}\left(P_{\mu\mu}^{\alpha}P_{\lambda\lambda}^{\alpha} + P_{\mu\mu}^{\beta}P_{\lambda\lambda}^{\beta}\right)(\mu\lambda|\lambda\mu). \tag{98}$$

Evidently, at large separations, the exchange energy consists of monoatomic and pairwise-additive diatomic terms:

$$E_x|_{HF}^{R\gg 0} = \sum_{a} E_x^a + \sum_{b>a} E_{x,ba}^{R\gg 0}. \tag{99}$$

*Pairwise additive inter-atomic correlation.* When discussing the asymptotic pairwise additivity of the correlation energy, interactions involving non-degenerate and degenerate atomic states must be distinguished. According to the Mandelstam-Tamm time-energy uncertainty relation $\Delta E \Delta t \sim \hbar$,[241] for any finite time $\Delta t$ of observing the system, the uncertainty $\Delta E$ in energy values of atomic states is greater than zero. Since for an arbitrary large $R$, inter-atomic integrals can be made arbitrary small and thus smaller than $\Delta E$, it is sufficient to consider the pairwise additivity in the case of non-degenerate atomic states.

If all atomic states are non-degenerate, at large separations the delocalized HF orbitals will transform to atomic orbitals. Leading second order terms would take the form

$$\frac{1}{4}\sum_{abrs}\frac{|(ar|bs)|^2}{\varepsilon_a + \varepsilon_b - \varepsilon_r - \varepsilon_s}, \tag{100}$$

where all $a$, $r$, $b$, and $s$ orbitals are localized on atoms. As such and other terms are additive in the perturbation theory, the asymptotic correlation energy up to all orders in PT can be written in terms of one-body, two-body, etc. interactions:

$$E_c|_{CI}^{R\gg 0} = \sum_{a} E_c^a + \sum_{a<b} E_c^{ab}$$
$$+ \sum_{a<b<c} E_c^{abc} + \cdots, \tag{101}$$

It is evident that to the lowest order in the inter-atomic overlap, the inter-atomic correlation energy is pairwise-additive.

**Supporting Information for**

# Self-Consistent Equations for Nonempirical Tight Binding Theory

Alexander V. Mironenko
*Department of Chemical and Biomolecular Engineering,*
The University of Illinois Urbana-Champaign, Urbana, Illinois 61801

**Section S1.**
Herein, it is proven that $\partial E_{xc}[\rho]/\partial q_{ak} = \langle \varphi_{ak}|\mu_{xc}|\varphi_{ak}\rangle$. By definition of the partial derivative,

$$\frac{\partial E_{xc}[\rho]}{\partial q_{ak}} = \lim_{\varepsilon\to 0} \frac{E_{xc}\begin{bmatrix}(q_{ak}+\varepsilon)\rho_{ak} \\ +\sum_{j\neq ak} q_j\rho_j\end{bmatrix} - E_{xc}[\rho]}{\varepsilon}. \tag{S1}$$

Since the functional derivative is defined as

$$\int \frac{\delta F[\rho]}{\delta \rho}\phi(\mathbf{r})d\mathbf{r} = \lim_{\varepsilon\to 0}\frac{F[\rho+\varepsilon\phi]-F[\rho]}{\varepsilon}, \tag{S2}$$

it follows from eq. (S1) that $F \equiv E_{xc}$, $\phi \equiv \rho_{ak}$, and thus

$$\frac{\partial E_{xc}[\rho]}{\partial q_{ak}} = \int \frac{\delta E_{xc}[\rho]}{\delta\rho}\rho_{ak}(\mathbf{r})d\mathbf{r}$$
$$= \int \frac{1}{q_{ak}}\frac{\delta E_{xc}[\rho]}{\delta\rho_{ak}}\rho_{ak}(\mathbf{r})d\mathbf{r} = \langle \varphi_{ak}|\mu_{xc}|\varphi_{ak}\rangle, \tag{S3}$$

where eq. (38) of the main manuscript was used, $\delta\rho$ was recognized to be any perturbation that preserves the number of electrons, and $\delta\rho = q_{ak}\delta\rho_{ak}$ was taken. QED.

**Section S2. Connection between the highest atomions and chemical potential of electrons.**
In this Section, it is shown that energy of the valence (highest-energy) atomion is associated with the electron chemical potential of a molecule. Chemical potential of electrons is, by definition,[1]

$$\mu = \frac{\partial E}{\partial N}. \tag{S4}$$

Yang et al. have proven[2] that, in the KS-DFT theory, $E[\rho] = E[v_s, N]$, where $v_s = \sum_a v_a(\mathbf{r}) + \phi(\mathbf{r}) + \mu_{xc}$ is the one-electron KS potential that defines the ground-state density $\rho$. In the independent atom ansatz, it is the set of potentials $v_{HA}^{ak}$ that defines the same density (eq. (24) of the main text). Evidently, the arguments by Yang et al. can be applied to NTB with no modification to show that $E[\rho] = E\left[\{v_{HA}^{ak}\}, N\right]$ and $E = \min_{v_{HA}^{ak}} E\left[\{v_{HA}^{ak}\}, N\right]$.

Consequently,[3]

$$\frac{\partial E}{\partial N} = \left(\frac{\partial E}{\partial N}\right)_{\{v_{HA}^{ak}\}}. \tag{S5}$$

Applying the argument due to Cohen et al.[3] to the independent atom ansatz, we find that

$$\left(\frac{\partial E}{\partial N}\right)_{\{v_{HA}^{ak}\}} = \varepsilon, \tag{S6}$$

where $\varepsilon$ is the highest occupied atomion energy for the degenerate atom-localized states, according to the atomion equalization principle (see Section IVC). Therefore,

$$\mu = \varepsilon, \tag{S7}$$

and the highest energy of occupied atomions equals the chemical potential of electrons in the system.

**Section S3. Origin of overlap-dependent prefactors.**

In this Section, the overlap-dependent prefactors, reported in Table 1 of the main manuscript for $H_2$, are derived. The Table is reproduced here:

| | $S$-dependent prefactor | Lowest order in $S$ |
|---|---|---|
| HF/CI | $(1+S)^{-2}$ | $-2S$ |
| VB | $(1+S^2)^{-1}$ | $-S^2$ |
| KS-DFT | $(1+S)^{-1}$ | $-S$ |
| 1-particle CI | $(1+S)^{-1}$ | $-S$ |

The last column entries are obtained by Taylor-expanding expressions in the second column to the lowest order in overlap. The prefactor for HF/CI in the second column is obtained by starting with the inter-atomic X energy for the diagonal density matrix, formulated in Appendix A:

$$E_x^{\circ\circ}|_{HF/CI}^{R\gg 0} = -\frac{1}{2}\sum_{\mu\lambda}\left(P_{\mu\mu}^{\alpha}P_{\lambda\lambda}^{\alpha} + P_{\mu\mu}^{\beta}P_{\lambda\lambda}^{\beta}\right)(\mu\lambda|\lambda\mu). \tag{S8}$$

We recognize that $2P_{\mu\mu}^{\alpha}P_{\lambda\lambda}^{\alpha} = 2c_\mu^2 c_\lambda^2$, where $c_\mu = c_\lambda = 1/\sqrt{2}$ are the orbital expansion coefficients. In the presence of overlap, they become $1/\sqrt{2(1+S)}$, which would yield an additional $1/(1+S)^2$ prefactor in front of $-0.5(\mu\lambda|\lambda\mu)$ inter-atomic terms. The VB prefactor follows from eq. (104) of the main manuscript. The KS-DFT and 1-particle CI prefactors follow from the fact that in the minimal-basis molecular orbital theory, which is isomorphic with the latter[4] and also with KS-DFT, the energy of the $H_2$ molecule is[5]

$$E_{MO} = 2\frac{H_{11} + H_{12}}{1 + S}, \tag{S9}$$

where $H_{ij} = \langle \varphi_i | H | \varphi_j \rangle$.

**Section S4. Derivation of the atomion equation.**

In this section, we derive the atomion equation from the asymptotic correspondence principle by generalizing the procedure reported in Paper I. We begin by summarizing the KS-DFT theory. In this theory, the electron density is computed as

$$\rho(\mathbf{r}) = \sum_i f_i \rho_i(\mathbf{r}), \tag{S10}$$

where $f_i$ is the occupancy of the KS state $|\psi_i\rangle$ and $\rho_i(\mathbf{r})$ is its density:

$$\rho_i(\mathbf{r}) = |\psi_i(\mathbf{r})|^2 \tag{S11}$$

The $|\psi_i\rangle$ states are the solutions to the KS equations:

$$\begin{gathered} H_{KS}|\psi_i\rangle = \varepsilon_i |\psi_i\rangle, \\ H_{KS} = -\frac{1}{2}\nabla^2 + \sum_a v_a(\mathbf{r}) \\ +\phi(\mathbf{r}) + \mu_{xc}, \end{gathered} \tag{S12}$$

where $H_{KS}$ is the KS Hamiltonian operator, $\mu_{xc} = \partial E_{xc}/\partial\rho$ is the XC potential, $\phi(\mathbf{r}) = \int \rho(\mathbf{r}') \frac{1}{|\mathbf{r}-\mathbf{r}'|} d\mathbf{r}'$ is the Hartree potential, and $v_a(\mathbf{r})$ is the potential due to the nucleus *a*.

The total KS energy functional has the form

$$\begin{gathered} E[\rho] = \sum_i f_i \varepsilon_i - \int \rho(\mathbf{r})\, \mu_{xc}[\rho] d\mathbf{r} \\ -\frac{1}{2}\int \rho(\mathbf{r})\, \phi[\rho] d\mathbf{r} + E_{xc}[\rho] + E_{NN}, \end{gathered} \tag{S13}$$

where $\varepsilon_i$ are one-electron KS energies. Herein, we assume that the exact XC functional is known, so that $E$, $\rho$, $\{\varepsilon_i\}$, and $\{\psi_i\}$ are also formally exact.

Next, we expand $|\psi_i\rangle$ states in a complete basis set of atom-localized states $|\varphi_{ak}\rangle$, where any $|\varphi_{ak}\rangle$ and $|\varphi_{al}\rangle$ states belonging to the same atom $a$ are taken as orthogonal. The expansion $|\psi_i\rangle = \sum_a c_{ai}|\varphi_a\rangle$ converts eq. (S12) into the generalized matrix eigenvalue problem – *the Kohn-Sham-Roothaan (KSR) equations*:

$$\mathbf{H}_{KS}\boldsymbol{c}_i = \varepsilon_i \mathbf{S}\boldsymbol{c}_i, \tag{S14}$$

where $\mathbf{c}_i = \{c_{ai}\}$ is an eigenvector for a KS eigenvalue $\varepsilon_i$ (distinct from the atomion energy $\varepsilon_{ak}$), and

$$\begin{gathered} (\mathbf{H}_{KS})_{ab} = H_{ab} = \langle \varphi_a | H_{KS} | \varphi_b \rangle, \\ (\mathbf{S})_{ab} = S_{ab} = \langle \varphi_a | \varphi_b \rangle. \end{gathered} \tag{S15}$$

Note that the definition of $\mathbf{S}$ differs from eq. (9) in the main manuscript. In eq. (S15) and hereafter the one-orbital convention is employed for simplicity – every atom is assumed to contribute only one atomion or atomic orbital to the system, unless stated overwise. This convention is exact for systems of 1- and 2-electron atoms in the ground state. For more complicated atoms and molecules, $a$ and $b$ indices are to be replaced with $ak$ and $be$, respectively, and summations over $k$ and $e$ atomic orbital indices are to be added. Next, we multiply both sides of eq. (S14) by the inverse overlap matrix $(\mathbf{S})^{-1}$ on the left to obtain

$$\mathbf{G}\boldsymbol{c}_i = \varepsilon_i \boldsymbol{c}_i,$$
$$\mathbf{G} = (\mathbf{S})^{-1}\mathbf{H}_{KS}. \tag{S16}$$

When the atoms are far apart, overlap matrix elements $S_{ab}$ are small. Applying the Löwdin expansion[6, 7] to $(\mathbf{S})^{-1}$ in the asymptotic limit of large inter-atomic separations, we find that, to the first order in the inter-atomic overlap $S_{ab}$,

$$((\mathbf{S})^{-1})_{ab} = 2\delta_{ab} - S_{ab} + O(S^2). \tag{S17}$$

The diagonal elements of the $\mathbf{G}$ matrix then become

$$G_{aa} = H_{aa} - \sum_{b \neq a} S_{ab} H_{ba}\,, \tag{S18}$$

In the same $O(S)$ limit, every atom can be regarded as weakly interacting with the environment and is thus independent. This enables the application of the variational principle to individual atoms – as opposed to the entire system. Formally, $\langle \varphi_a | \delta\varphi_b \rangle \ll \langle \varphi_a | \varphi_b \rangle$, and for nearly constant $\varphi_a$ in the vicinity of $\varphi_b$ at $r \gg 0$, $\langle \delta\varphi_b | H_{KS} | \varphi_a \rangle \sim \langle \delta\varphi_b | \varphi_a \rangle$, so that $\delta\varphi_b$ variations have a negligible effect on $G_{aa}$, unlike $\delta\varphi_a$. Variational optimization of eq. (S18) under the $\varphi_a$ normalization constraint leads to the following eigenvalue problem:

$$\varepsilon_a |\varphi_a\rangle = \left[ H_{KS} - \sum_{b \neq a} P_b H_{KS} \right] |\varphi_a\rangle, \tag{S19}$$

where the projection operator $P_b = |\varphi_b\rangle\langle\varphi_b|$, and $\varepsilon_a = G_{aa}$ is regarded as the effective energy of the state $|\varphi_a\rangle$. Eq. (S19), derived from the KS theory in the $O(S)$ limit, shall be asymptotically correspondent to the atomion eigenvalue problem in the independent atom ansatz, and the $|\varphi_a\rangle$ states shall be identical to atomions.

To relate $H_{KS}$ in eq. (S19) to $H_{HA}^{a}$ in the independent atom ansatz (see eq. (24) of the main text), it is noted that to $O(S)$, $\rho \to \sum_b \rho_b$, $\phi \to \sum_b \phi_b$, and

$$\mu_{xc}|\varphi_a\rangle = \frac{\partial E_{xc}}{\partial \rho}\bigg|\varphi_a\bigg\rangle \to \frac{\partial E_{xc}}{\partial \rho_a}\bigg|\varphi_a\bigg\rangle, \tag{S20}$$

since $\left\langle \varphi_a \middle| \frac{\partial E_{xc}}{\partial \rho} \middle| \varphi_a \right\rangle$ is equivalent to the functional derivative containing $E_{xc}[\rho + \varepsilon\rho_a]$, which is the same as $E_{xc}[\rho_a + \varepsilon\rho_a, \{\rho_b | b \neq a\}]$ for $\rho = \sum_b \rho_b$. In Section VC, it was found that at $R \gg 0$, $E_{xc}$ is split into intra-atomic and inter-atomic parts, where the latter can be shown to be pairwise-additive to $O(S)$. Asymptotic pairwise additivity is discussed in more detail in Appendix C. It follows from it that the asymptotic form of the XC potential acting on $|\varphi_a\rangle$ is the sum of intra-atomic and pairwise-additive inter-atomic terms:

$$\mu_{xc}|\varphi_a\rangle \to \mu_{xc}^a|\varphi_a\rangle + \sum_{b\neq a} \mu_{xc,ba}^{R\gg 0}|\varphi_a\rangle, \tag{S21}$$

where $\mu_{xc}^a$ was defined in eq. (38), and $\mu_{xc,ba}^{R\gg 0}$ is the inter-atomic XC potential involving only two atoms: $a$ and $b$. In the asymptotic limit, we thus have

$$H_{KS}|\varphi_a\rangle \to H_{KS,a}^{R\gg 0}|\varphi_a\rangle, \tag{S22}$$

where

$$H_{KS,a}^{R\gg 0}|\varphi_a\rangle = \left(-\frac{1}{2}\nabla^2 + \sum_b v_{es}^b + \mu_{xc}^a + \sum_{b\neq a} \mu_{xc,ba}^{R\gg 0}\right)|\varphi_a\rangle. \tag{S23}$$

In eq. (S23), the $v_{es}^b$ quantity was defined in eq. (36). Eq. (S19) can then be written as

$$\varepsilon_a|\varphi_a\rangle = \left[H_{KS,a}^{R\gg 0} - \sum_{b\neq a} P_b H_{KS',a}^{R\gg 0}\right]|\varphi_a\rangle,$$
$$H_{KS',a}^{R\gg 0} = H_{KS,a}^{R\gg 0} \tag{S24}$$

It can be shown that, since $H_{KS,a}^{R\gg 0}$ appears twice in eq. (S24), the expression is strictly valid in the limit $S \to 0$, $R \to \infty$, and does not hold for any non-zero $S$ (Section S5). To rework eq. (S24) into a more general form applicable for $S > 0$, it is first noted that in the united atom limit with $R_{ab} \to 0$ and $S \to 1$, $\varepsilon_a \to \langle\varphi_a|H_{KS,a}^{R\gg 0} - H_{KS',a}^{R\gg 0}|\varphi_a\rangle$. Since an atomic Hamiltonian, and thus $H_{KS,a}^{R\gg 0} - H_{KS',a}^{R\gg 0}$, must contain kinetic and electrostatic operators and intra-atomic XC potentials, $H_{KS',a}^{R\gg 0}$ cannot include these terms to avoid their unphysical cancellation. To eliminate such terms from $H_{KS',a}^{R\gg 0}$ in a systematic manner, while maintaining consistency with eq. (S24), it is first recognized that at $S \to 0$, the $S$-dependent terms have a first-order effect on $\varepsilon_a$ and thus do not affect $|\varphi_a\rangle$. The $|\varphi_a\rangle$ state can then be regarded as an eigenfunction of the $S$-independent Hamiltonian:

$$\varepsilon_a^0|\varphi_a\rangle = \left[-\frac{1}{2}\nabla^2 + \sum_b v_{es}^b \right. \\ \left. +\mu_{xc}^a\right]|\varphi_a\rangle. \tag{S25}$$

Among many $|\varphi_a\rangle$ and $\varepsilon_a^0$ solutions, partially occupied valence atomions are degenerate according to the atomion equalization principle (Section IVC):

$$\varepsilon_a^0 = \varepsilon_b^0 = \cdots = \varepsilon^0. \tag{S26}$$

The degenerate states make the $\sim H_{ab}$ contributions to molecular orbital (MO) energies $\varepsilon_i$, whereas the non-degenerate states (e.g., $s$ and $p$ orbitals on different atoms) contribute $\sim H_{ab}^2/|\varepsilon_a - \varepsilon_b|$ at $R \gg 0$. Since the interatomic integrals must become local in this limit, according to Section VII, the degenerate state contributions are of $O(\varphi_a\varphi_b)$, while the non-degenerate contributions are of $O((\varphi_a\varphi_b)^2)$. Thus, to the lowest $O(\varphi_a\varphi_b)$, relevant for the $S \to 0$ limit considered here, MOs can be regarded as the superpositions of only degenerate valence atomions.

For valence atomions, substitution of eq. (S25) with $\varepsilon_a^0 = \varepsilon^0$ into $H_{KS',a}^{R\gg 0}|\varphi_a\rangle$ in eq. (S24), where $H_{KS',a}^{R\gg 0} = H_{KS,a}^{R\gg 0}$ is defined in eq. (S23), leads to

$$\varepsilon_a|\varphi_a\rangle = \left[H_{KS,a}^{R\gg 0} \right. \\ \left. -\sum_{b\neq a} P_b\left(\varepsilon^0 + \sum_{c\neq a}\mu_{xc,ca}^{R\gg 0}\right)\right]|\varphi_a\rangle. \tag{S27}$$

The constant shift $H_{KS} \to H_{KS} - \varepsilon^0$ results in all KS energies shifted by the same amount: $\varepsilon_i \to \varepsilon_i - \varepsilon^0$. This shift can be compensated by adding $\varepsilon^0$ to the diagonal elements of the **G** matrix in eq. (S16), so that

$$(\mathbf{G} + \varepsilon^0\mathbf{I})\boldsymbol{c}_i = \varepsilon_i\boldsymbol{c}_i. \tag{S28}$$

Repeating the derivation of eq. (S24) using the shifted $H_{KS}$, we find that

$$\varepsilon_a|\varphi_a\rangle = \left[H_{KS,a}^{R\gg 0} \right. \\ \left. -\sum_{b\neq a} P_b\left(H_{KS',a}^{R\gg 0} - \varepsilon^0\right)\right]|\varphi_a\rangle \tag{S29}$$

After recognizing that $H_{KS'}^{\infty,a} - \varepsilon^0 = \sum_{c\neq a} \mu_{xc,ca}^{R\gg 0}$ (cf. eq. (S27)), eq. (S29) becomes

$$\varepsilon_a|\varphi_a\rangle = \left[H_{KS,a}^{R\gg 0} - \sum_{b\neq a} P_b\left(\sum_{c\neq a}\mu_{xc,ca}^{R\gg 0}\right)\right]|\varphi_a\rangle. \tag{S30}$$

Terms in parentheses of eq. (SS30) can be classified as involving either two-body $\langle\varphi_b|\mu_{xc,ba}^{R\gg 0}|\varphi_a\rangle$ or three-body $\langle\varphi_b|\mu_{xc,ca}^{R\gg 0}|\varphi_a\rangle$ integrals. From the locality of inter-atomic integrals in the large separation limit (Section VII) it follows that $\langle\varphi_b|\mu_{xc,ba}^{R\gg 0}|\varphi_a\rangle \sim S_{ba}$ and $\langle\varphi_b|\mu_{xc,ca}^{R\gg 0}|\varphi_a\rangle \sim S_{ba}S_{ca}S_{bc}$, so that to the lowest $O(\varphi_a\varphi_b)$, the 3-body terms can be neglected to yield

$$\varepsilon_a|\varphi_a\rangle = \left[H_{KS,a}^{R\gg 0} - \sum_{b\neq a} P_b\mu_{xc,ba}^{R\gg 0}\right]|\varphi_a\rangle. \tag{S31}$$

Finally, after substituting $H_{KS,a}^{R\gg 0}$ from eq. (S23) into eq. (S31), the following *atomion equation* is obtained:

$$\varepsilon_a|\varphi_a\rangle = \Bigg[-\frac{1}{2}\nabla^2 + v_{es}^a + \mu_{xc}^a + \sum_{b\neq a}\left(v_{es}^b + \mu_{xc,ba}^{R\gg 0} - P_b\mu_{xc,ba}^{R\gg 0}\right)\Bigg]|\varphi_a\rangle. \tag{S32}$$

**Section S5.**

In this Section, I show that the following equation only holds in the limit $S \to 0$:

$$\varepsilon_a|\varphi_a\rangle = \left[H_{KS}^{R\gg 0,a} - \sum_{b\neq a} P_b H_{KS'}^{R\gg 0,a}\right]|\varphi_a\rangle,$$
$$H_{KS'}^{R\gg 0,a} = H_{KS}^{R\gg 0,a} \tag{S33}$$

The equation can be written as

$$H_{KS}^{R\gg 0,a}|\varphi_a\rangle = \varepsilon_a|\varphi_a\rangle + \sum_{b\neq a}|\varphi_b\rangle\langle\varphi_b|H_{KS}^{R\gg 0,a}|\varphi_a\rangle. \tag{S34}$$

After substituting $H_{KS}^{R\gg 0,a}|\varphi_a\rangle$ of eq. (S34) into $P_b H_{KS}^{R\gg 0,a}|\varphi_a\rangle$ of eq. (S33) and rearranging, it follows that

$$H_{KS}^{R\gg 0,a}|\varphi_a\rangle = \varepsilon_a|\varphi_a\rangle + \sum_{b\neq a}|\varphi_b\rangle\langle\varphi_b|H_{KS}^{R\gg 0,a}|\varphi_a\rangle +$$

$$\varepsilon_a \sum_{b\neq a} S_{ba}\,|\varphi_b\rangle +$$

$$\sum_{b\neq a}\sum_{\substack{b'\neq a',\\ b'\neq b}} |\varphi_b\rangle S_{bb'}\,\langle\varphi_{b'}|H_{KS}^{R\gg 0,a}|\varphi_a\rangle\,, \tag{S35}$$

Evidently, eq. (S34) and eq. (S35) are simultaneously valid only when $S = 0$.

**Section S6. Derivation of Hückel eigenproblem**

In this section, Hückel theory is obtained from the atomion equation. The approach is essentially identical to that presented in Paper I, with a slightly different notation.

To unravel the eigenvalue problem that gives rise to bonding/antibonding contributions to $E_{xc}^{\infty}$, eq. (S32) is re-written as

$$H_{KS,a}^{R\gg 0}|\varphi_a\rangle = D_{aa}|\varphi_a\rangle + \sum_{b\neq a} D_{ba}|\varphi_b\rangle\,, \tag{S36}$$

where

$$D_{ba} = \langle\varphi_b|\mu_{xc,ba}^{R\gg 0}|\varphi_a\rangle \text{ for } b\neq a \tag{S37}$$

$$D_{aa} = \varepsilon_a = H_{aa} - \sum_{b\neq a} S_{ab}D_{ba}\,, \tag{S38}$$

$$H_{aa} = \langle\varphi_a|H_{KS,a}^{R\gg 0}|\varphi_a\rangle, \tag{S39}$$

and where $H_{KS,a}^{R\gg 0}$ (the asymptotic form of $H_{KS}$) is defined in eq. (S23). For the KS $|\psi_i\rangle$ states, expanded in the basis set of $|\varphi_a\rangle$ states, in the asymptotic limit we have

$$H_{KS}|\psi_i\rangle = \sum_a c_{ai}\,H_{KS}|\varphi_a\rangle$$

$$\to \sum_a c_{ai}\,H_{KS,a}^{R\gg 0}|\varphi_a\rangle, \tag{S40}$$

according to eq. (S22). After substituting eq. (S36) into the right-hand side of eq. (S40), recognizing that $H_{KS}|\psi_i\rangle = \varepsilon_i^{mo}|\psi_i\rangle$, and further expanding $|\psi_i\rangle$ in the $|\varphi_a\rangle$ basis set, one obtains

$$\sum_{a,b}(c_{ai}\varepsilon_i^{mo} - c_{bi}D_{ab})\,|\varphi_a\rangle = 0, \tag{S41}$$

Since the $|\varphi_a\rangle$ states are linearly independent, the secular determinant of eq. (S41) is zero. Standard eigenvalue problem then follows:

$$\mathbf{D}\mathbf{c}_i = \varepsilon_i^{mo}\mathbf{c}_i, \tag{S42}$$

where the **D** matrix contains $D_{aa}$ and $D_{ba}$ elements defined in eq. (S37)-(S39). Eq. (S42) is identical to the Hückel (orthogonal tight-binding) electronic structure problem,[8] but with non-empirical **D** matrix elements. Throughout the article, $\varepsilon_i^{mo}$ are referred to as either Hückel energies or molecular orbital (MO) energies, and off-diagonal $D_{ab}$ – as resonance integrals.

**Section S7. Derivation of the total energy expression.**

In this Section, the NTB total energy expression is derived. Since the Hückel states normalize according to $\mathbf{c}_i^T\mathbf{c}_i = 1$, they are consistent with the atomic density additivity and the independent atom ansatz (Section III). The total Hückel energy is

$$\sum_i f_i\varepsilon_i^{mo} = \sum_a o_a\varepsilon_a + \sum_{a,b\neq a} p_{ab}D_{ab}\,, \tag{S43}$$

where the Hückel charge $o_a$ is defined as

$$o_a = \sum_i f_i c_{ai}^2\,, \tag{S44}$$

and the bond order $p_{ab}$ is

$$p_{ab} = \sum_i f_i c_{ai}^* c_{bi}\,. \tag{S45}$$

In the asymptotic limit, only the degenerate states $\varepsilon_a$ mix in eq. (S43) (see Section S4), so that $\sum_a o_a\varepsilon_a$ can be replaced with $\sum_a q_a\varepsilon_a$ with no loss in accuracy, where $q_a$ was defined in eq. (18), since $\sum_a q_a = \sum_a o_a = N_v$ and $N_v$ is the number of valence electrons in the molecule. The replacement procedure is formally equivalent to rescaling $o_a$ by some factors and would imply that there are two simultaneous mechanisms of charge transfer in the NTB theory – through electronegativity equilibration and orbital hybridization.

To proceed, eq. (S13) of KS-DFT is written to $O(S)$ by employing the electron density additivity (eq. (17) and (18)) and the pairwise-additive form of $\mu_{xc}$ (eq. (S21)) to get ($\rho_a$ is redefined as $|\varphi_a|^2$):

$$E[\rho] = \sum_i f_i \varepsilon_i^{mo}$$
$$-\sum_a q_a \int \rho_a(\mathbf{r}) \left( \mu_{xc}^a + \sum_{b \neq a} \mu_{xc,ba}^{R \gg 0} \right) d\mathbf{r}$$
$$-\frac{1}{2} \sum_{a,b} q_a q_b \int \frac{\rho_b(\mathbf{r}) \rho_a(\mathbf{r}')}{|\mathbf{r} - \mathbf{r}'|} d\mathbf{r} d\mathbf{r}'$$
$$+E_{xc}\left[\sum_a q_a \rho_a\right] + E_{NN}, \quad \text{(S46)}$$

Since the Hückel eigenvalues coincide with the KS eigenvalues as has been shown in Section VE, eq. (S43) can be substituted into eq. (S46) after $o_a \rightarrow q_a$ replacement, using eq. (S38), (S39), and (S23). After cancellation of the terms containing $\mu_{xc}$, the following NTB total energy expression is obtained for combinations of one-electron atoms:

$$E = \sum_a q_a \left\langle \varphi_a \middle| -\frac{1}{2} \nabla^2 \middle| \varphi_a \right\rangle$$
$$- \sum_{a, b \neq a} q_a S_{ab} \langle \varphi_b | \mu_{xc,ba}^{R \gg 0} | \varphi_a \rangle$$
$$+ \sum_{a, b \neq a} p_{ab} \langle \varphi_a | \mu_{xc,ab}^{R \gg 0} | \varphi_b \rangle$$
$$+E_{es}\left[\sum_a q_a \rho_a\right] + E_{xc}\left[\sum_a q_a \rho_a\right], \quad \text{(S47)}$$

where $E_{es}$ was defined in eq. (30).

**Section S8. Proof of the force theorem in the NTB theory**

A tiny displacement of a nucleus $a$ in a molecule causes energy change due to its explicit and implicit dependence on nuclear coordinates $\mathbf{R}_a$:

$$\frac{dE}{d\mathbf{R}_a} = \left(\frac{\partial E}{\partial \mathbf{R}_I}\right)_{\rho,N} + \int \left(\frac{\delta E}{\delta \rho(\mathbf{r})}\right)_{R_a,N} \frac{\partial \rho(\mathbf{r})}{\partial \mathbf{R}_a} d\mathbf{r} \quad \text{(S48)}$$

Since $\left(\frac{\delta E}{\delta \rho(\mathbf{r})}\right)_{R_a,N} = 0$ according to the DFT variational principle, only the explicit $E$ dependence contributes to the total derivative:

$$\frac{dE}{d\mathbf{R}_a} = \left(\frac{\partial E}{\partial \mathbf{R}_a}\right)_{\rho,N} = \int \rho(\mathbf{r}) \frac{\partial v_{ext}}{\partial \mathbf{R}_a} d\mathbf{r} + \frac{\partial E_{NN}}{\partial \mathbf{R}_a} \quad \text{(S49)}$$

This is the famous Hellmann-Feynman (electrostatic) force theorem[9, 10] – the total derivative of energy with respect to a nuclear displacement contains only electrostatic contributions. It can be interpreted as if the nucleus displacement does not move density with it, and thus no density-dependent energy changes arise.

As the NTB theory is formally exact, its total energy expression must satisfy the force theorem. Intuitively, the atomion equation and the Hückel eigenvalue problem, derived through application of the variational principle, suggest that both expansion coefficients and atomion shapes shall remain frozen upon nucleus displacement. Thus, the electron density remains frozen and the force theorem is satisfied.

To prove the NTB force theorem, I first write the NTB total energy expression as a sum of the "atomion superposition" $E_{sup} = E_{kin} + E_{ortho} + E_{xc} + E_{es}$ and interatomic mixing $E_{hyb}$ energy terms:

$$E = E_{sup} + E_{hyb} \tag{S50}$$

For $E = E[\{\mathbf{R}_a\}, \{|\varphi_a|^2\}, \{q_a\}, \{c_a\}]$, the energy derivative with respect to the displacement becomes

$$\begin{aligned}\frac{\partial E}{\partial \mathbf{R}_a} = \int \rho(\mathbf{r}) \frac{\partial v_{ext}}{\partial \mathbf{R}_a} d\mathbf{r} + \frac{\partial E_{NN}}{\partial \mathbf{R}_a} + \sum_{a \neq b} p_{ab} \left\langle \varphi_a \middle| \frac{\partial \mu_{xc,ab}^{R\gg 0}}{\partial \mathbf{R}_a} \middle| \varphi_b \right\rangle - \sum_{a \neq b} q_a S_{ab} \left\langle \varphi_b \middle| \frac{\partial \mu_{xc,ba}^{R\gg 0}}{\partial \mathbf{R}_a} \middle| \varphi_a \right\rangle \\ + \sum_a \int \frac{\delta E_{sup}}{\delta |\varphi_a(\mathbf{r})|^2} \frac{\partial |\varphi_a(\mathbf{r})|^2}{\partial \mathbf{R}_a} d\mathbf{r} + \sum_a \int \frac{\delta E_{hyb}}{\delta |\varphi_a(\mathbf{r})|^2} \frac{\partial |\varphi_a(\mathbf{r})|^2}{\partial \mathbf{R}_a} d\mathbf{r} \\ + \sum_a \int \frac{\delta E}{\delta c_a} \frac{\partial c_a}{\partial \mathbf{R}_a} d\mathbf{r} + \sum_a \int \frac{\delta E}{\delta q_a} \frac{\partial q_a}{\partial \mathbf{R}_a} d\mathbf{r},\end{aligned} \tag{S51}$$

To satisfy the force theorem, all the terms except the first two must be equal to zero. As the $\mu_{xc,ab}^{R\gg 0}$ operator does not explicitly depend on nuclear potentials (see Section VII), $\frac{\partial \mu_{xc,ab}^{R\gg 0}}{\partial \mathbf{R}_a} = \frac{\partial \mu_{xc,ba}^{R\gg 0}}{\partial \mathbf{R}_a} = 0$. $\frac{\delta E_{sup}}{\delta |\varphi_a(\mathbf{r})|^2} = 0$ is equivalent to the NTB variational principle that leads to the atomion equation, whereas $\frac{\delta E_{hyb}}{\delta |\varphi_a(\mathbf{r})|^2} = 0$ since $E_{hyb}$ is regarded as not being a functional of density. $\frac{\delta E}{\delta c_a} = 0$ as a consequence of the Hückel problem, and $\frac{\delta E}{\delta q_a} = 0$ as a consequence of the atomion equalization principle. The NTB force theorem is proved.

## Section S9.

To obtain the local form of eq. (86), a spherical coordinate system is adopted with the origin placed on nucleus $b$ and the z axis pointing toward nucleus $a$. The infinitesimal volume element $d\mathbf{r}'$ is then $(r')^2 dr' \sin\theta' \, d\theta' d\varphi'$. The $1s$ orbital $\varphi_b^*(\mathbf{r}')$ is a product of radial and angular parts:

$$\varphi_b^*(r') = Y_0^{0*}(\theta', \varphi') R_b^*(r') = \frac{1}{2\sqrt{\pi}} R_b^*(r'), \tag{S52}$$

and the $1/|\mathbf{r} - \mathbf{r}'|$ potential has the following Laplace expansion:

$$\frac{1}{|\mathbf{r}-\mathbf{r}'|} = \sum_{L=0}^{\infty} \frac{4\pi}{2L+1}$$

$$\sum_{m=-L}^{+L} \frac{r_<^L}{r_>^{L+1}} Y_L^{m*}(\theta,\varphi) Y_L^m(\theta',\varphi'), \tag{S53}$$

where $r_< = \min(r, r')$ and $r_> = \max(r, r')$ and where the $Y_L^{m*}(\theta,\varphi) = (-1)^m Y_L^{-m}(\theta,\varphi)$ substitution was made.

In the $R \gg 0$ limit, $\varphi_a$ can be replaced with a constant $\varphi_a^\infty$ near the origin where the $\varphi_b$ magnitude is substantial. From eq. (86), (S52), and (S53) it follows that

$$\mu_{xc,ba}^{R\gg 0}(\mathbf{r}) = -\frac{1}{4}\varphi_b(\mathbf{r}) \sum_{L=0}^{\infty} \frac{4\pi}{2L+1} \sum_{m=-L}^{+L} \int \square$$

$$\frac{r_<^L}{r_>^{L+1}} Y_L^{m*}(\theta,\varphi) Y_L^m(\theta',\varphi') Y_0^{0*}(\theta',\varphi')$$

$$R_b^*(r')(r')^2 dr' \sin\theta' \, d\theta' d\varphi' \tag{S54}$$

Since $Y_L^m(\theta',\varphi')$ and $Y_0^{0*}(\theta',\varphi')$ are orthonormal, integral of their product over angular coordinates equals zero for $L \neq 0$ and unity for $L = 0$. Therefore,

$$\mu_{xc,ba}^{R\gg 0}(\mathbf{r}) = -\frac{1}{2}\sqrt{\pi}\varphi_b(\mathbf{r}) \int \frac{1}{r_>} R_b^*(r')(r')^2 dr'. \tag{S55}$$

**Supplemental References:**